\PassOptionsToPackage{usenames,dvipsnames}{xcolor}

\pdfoutput=1
\newcommand*{\ATLASLATEXPATH}{latex/}
\documentclass[cernpreprint,
UKenglish,
texlive=2016,
PAPER,
]{\ATLASLATEXPATH atlasdoc}
\usepackage[subcaption]{\ATLASLATEXPATH atlaspackage}
\usepackage{\ATLASLATEXPATH atlasbiblatex}

\usepackage{\ATLASLATEXPATH atlasphysics}
\usepackage[textsize=tiny]{todonotes}
\usepackage{multirow}
\usepackage{wrapfig}
\usepackage{floatrow}
\usepackage{makecell}
\usepackage{tabularx}

\usepackage{tikz}

\usetikzlibrary{arrows}
\usetikzlibrary{backgrounds}
\usetikzlibrary{calc}
\usetikzlibrary{decorations.text}
\usetikzlibrary{decorations.markings}
\usetikzlibrary{decorations.pathmorphing}
\usetikzlibrary{decorations.pathreplacing}
\usetikzlibrary{positioning}
\usetikzlibrary{shadows.blur}
\usetikzlibrary{shadows}
\usetikzlibrary{shapes}
\usetikzlibrary{shapes.symbols}
\usetikzlibrary{shapes.arrows}
\usetikzlibrary{intersections}
\usetikzlibrary{through}

\makeatletter
\pgfdeclaredecoration{gluon_decoration}{initial}
{
  \state{initial}[
    width=+0pt,
    next state=coil,
    persistent precomputation={
      \pgfmathsetmacro\matchinglength{
        (ceil(\pgfdecoratedinputsegmentlength / \pgfdecorationsegmentlength) - \pgfdecoratedinputsegmentlength / \pgfdecorationsegmentlength) > 0.5
        ? (\pgfdecoratedinputsegmentlength - 2 * \pgfdecorationsegmentaspect * \pgfdecorationsegmentamplitude) / (floor(\pgfdecoratedinputsegmentlength / \pgfdecorationsegmentlength) + 0.499)
        : (\pgfdecoratedinputsegmentlength - 2 * \pgfdecorationsegmentaspect * \pgfdecorationsegmentamplitude) / (ceil(\pgfdecoratedinputsegmentlength / \pgfdecorationsegmentlength) + 0.499)
      }
      \setlength{\pgfdecorationsegmentlength}{\matchinglength pt}
    },
  ]{}
  \state{coil}[
    switch if less than=%
      0.5\pgfdecorationsegmentlength%
      +\pgfdecorationsegmentaspect\pgfdecorationsegmentamplitude%
      +\pgfdecorationsegmentaspect\pgfdecorationsegmentamplitude to last,
    width=+\pgfdecorationsegmentlength,
  ]
  {
    \pgfpathcurveto
    {\pgfpoint@oncoil{0    }{ 0.555}{ 1}}
    {\pgfpoint@oncoil{0.445}{ 1    }{ 2}}
    {\pgfpoint@oncoil{1    }{ 1    }{ 3}}
    \pgfpathcurveto
    {\pgfpoint@oncoil{1.555}{ 1    }{ 4}}
    {\pgfpoint@oncoil{2    }{ 0.555}{ 5}}
    {\pgfpoint@oncoil{2    }{ 0    }{ 6}}
    \pgfpathcurveto
    {\pgfpoint@oncoil{2    }{-0.555}{ 7}}
    {\pgfpoint@oncoil{1.555}{-1    }{ 8}}
    {\pgfpoint@oncoil{1    }{-1    }{ 9}}
    \pgfpathcurveto
    {\pgfpoint@oncoil{0.445}{-1    }{10}}
    {\pgfpoint@oncoil{0    }{-0.555}{11}}
    {\pgfpoint@oncoil{0    }{ 0    }{12}}
  }
  \state{last}[next state=final]
  {
    \pgfpathcurveto
    {\pgfpoint@oncoil{0    }{ 0.555}{1}}
    {\pgfpoint@oncoil{0.445}{ 1    }{2}}
    {\pgfpoint@oncoil{1    }{ 1    }{3}}
    \pgfpathcurveto
    {\pgfpoint@oncoil{1.555}{ 1    }{4}}
    {\pgfpoint@oncoil{2    }{ 0.555}{5}}
    {\pgfpoint@oncoil{2    }{ 0    }{6}}
  }
  \state{final}{}
}

\makeatother

\graphicspath{{logos/}{figures/}}

\AtlasTitle{Measurement of colour flow using jet-pull observables in $t\bar{t}$ events with the ATLAS experiment at $\sqrt{s} = \SI{13}{\TeV}$}

\author{The ATLAS Collaboration}

\makeatletter
\ifADOC@PAPER
  \AtlasRefCode{TOPQ-2017-13}
  
  \else
    \ifADOC@CONF
      \AtlasNote{ATLAS-CONF-2017-069}
    \else
      \PackageError{customATLAS}{invalid document type}{document is neither paper nor conf note}
    \fi
\fi
\makeatother

\PreprintIdNumber{CERN-EP-2018-041}

\AtlasJournal{EPJC}
\AtlasAbstract{%
Previous studies have shown
that weighted angular moments derived from jet constituents encode the
colour connections between partons that seed the jets. This paper
presents measurements of two such distributions, the jet-pull angle and jet-pull
magnitude, both of which are derived from the jet-pull angular moment. The
measurement is performed in $t\bar{t}$ events with one leptonically decaying $W$~boson
and one hadronically decaying $W$~boson, using $36.1\,\text{fb}^{-1}$ of
$pp$ collision data recorded by the ATLAS detector at $\sqrt{s} = 13 \,
\text{TeV}$ delivered by the Large Hadron Collider. The observables are measured
for two dijet systems, corresponding to the colour-connected daughters of the
$W$~boson and the two $b$-jets from the top-quark decays.
To allow the comparison of the measured distributions to colour model predictions, the
measured distributions are unfolded to
particle level, after correcting for experimental effects introduced by the
detector. While good agreement can be found for some combinations of predictions
and observables, none of the predictions describes the data well across all
observables.
}

\AtlasCoverSupportingNote{Measurement of colour flow in $t\bar{t}$ events using the ATLAS detector at $\sqrt{s}=13$ TeV}{https://cds.cern.ch/record/2267944}
\AtlasCoverAnalysisTeam{Tom Neep, Yvonne Peters, Fabian Wilk}

\AtlasCoverEdBoardMember{Ariel~Schwartzman~(chair), Kenneth Johns, Ben Nachman}
\AtlasCoverEgroupEditors{atlas-topq-2017-13-editors@cern.ch}

\AtlasCoverEgroupEdBoard{atlas-topq-2017-13-editorial-board@cern.ch}

\hypersetup{pdftitle={ATLAS document},pdfauthor={The ATLAS Collaboration}}

\begin{document}

\maketitle

% \tableofcontents

% List of contributors - print here or after the Bibliography.
% \PrintAtlasContribute{0.30}
% \clearpage

% -------------------------------------------------------------------------------
\section{Introduction}
\label{sec:intro}
% -------------------------------------------------------------------------------

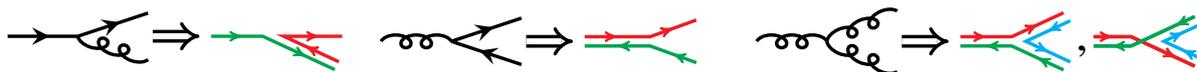
\begin{figure}[!htb]
  \centering
  % \selectcolormodel{gray}
  \begin{subfigure}{0.28\textwidth}
    \centering
    \scalebox{1.6}{\begin{tikzpicture}[thick]
  \tikzset{
    % style to apply some styles to each segment of a path
    on each segment/.style={
      decorate,
      decoration={
        show path construction,
        moveto code={},
        lineto code={
          \path [#1]
          (\tikzinputsegmentfirst) -- (\tikzinputsegmentlast);
        },
        curveto code={
          \path [#1] (\tikzinputsegmentfirst)
          .. controls
          (\tikzinputsegmentsupporta) and (\tikzinputsegmentsupportb)
          ..
          (\tikzinputsegmentlast);
        },
        closepath code={
          \path [#1]
          (\tikzinputsegmentfirst) -- (\tikzinputsegmentlast);
        },
      },
    },
    % style to add an arrow in the middle of a path
    mid arrow/.style={postaction={decorate,decoration={
          markings,
          mark=at position .6 with {\arrow[#1]{stealth}}
        }}},
    small mid arrow/.style={postaction={decorate,decoration={
          markings,
          mark=at position .6 with {\arrow[#1]{stealth'}}
        }}},
    boson_wdir/.style={draw=black, dashed, postaction={decorate}, decoration={markings,mark=at position .5 with {\arrow[draw=black]{>}}}},
    boson/.style={draw=black, dashed},
    photon/.style={decorate,decoration={snake},draw=black},
    gluon/.style={decorate,decoration={gluon_decoration, aspect=1.5, amplitude=1.5pt, segment length=8pt}},
    gluon_up/.style={decorate,decoration={gluon_decoration, mirror, aspect=1.5, amplitude=1.5pt, segment length=8pt}},
    fermion/.style={mid arrow}
  }
  
  \coordinate (feynman-q-in) at (0, 0);
  \coordinate (feynman-vtx) at ($(feynman-q-in) + (1.5em, 0em)$) ;
  \coordinate (feynman-q-out) at ($(feynman-vtx) + (1.5em, 0.5em)$);
  \coordinate (feynman-glu-out) at ($(feynman-vtx) + (1.5em, -0.5em)$);

  \coordinate (connector-lhs) at ($(feynman-vtx) + (1.6em, 0em)$);
  \coordinate (connector-rhs) at ($(connector-lhs) + (1em, 0em)$);

  \coordinate (cflow-q-in) at ($(connector-rhs) + (0.25em, 0em)$);
  \coordinate (cflow-vtx-closed) at ($(cflow-q-in) + (1.1em, 0em)$) ;
  \coordinate (cflow-vtx-open) at ($(cflow-vtx-closed) + (0.45em, 0em)$) ;

  \coordinate (cflow-q-out) at ($(cflow-vtx-open) + (1.25em, 0em)$);
  \coordinate (cflow-glu-out-top) at ($(cflow-vtx-open) + (-25:1.3em)$);
  \coordinate (cflow-glu-out-bot) at ($(cflow-vtx-closed) + (-25:1.7em)$);
  
  \draw[fermion] (feynman-q-in) -- (feynman-vtx);
  \draw[fermion] (feynman-vtx) -- (feynman-q-out);
  \draw[gluon] (feynman-glu-out) -- (feynman-vtx);

  \draw[-implies,double equal sign distance] (connector-lhs) -- (connector-rhs);

  \draw[Red, postaction={on each segment={small mid arrow={Red, scale=0.6}}}]
  (cflow-glu-out-top) -- (cflow-vtx-open) -- (cflow-q-out);
  
  \draw[Green, postaction={on each segment={small mid arrow={Green, scale=0.6}}}]
  (cflow-q-in) -- (cflow-vtx-closed) -- (cflow-glu-out-bot);
\end{tikzpicture}}
    % \rule{\columnwidth}{3pt}
  \end{subfigure}
  \hfill
  \begin{subfigure}{0.28\textwidth}
    \centering
    \scalebox{1.6}{\begin{tikzpicture}[thick]
  \tikzset{
    % style to apply some styles to each segment of a path
    on each segment/.style={
      decorate,
      decoration={
        show path construction,
        moveto code={},
        lineto code={
          \path [#1]
          (\tikzinputsegmentfirst) -- (\tikzinputsegmentlast);
        },
        curveto code={
          \path [#1] (\tikzinputsegmentfirst)
          .. controls
          (\tikzinputsegmentsupporta) and (\tikzinputsegmentsupportb)
          ..
          (\tikzinputsegmentlast);
        },
        closepath code={
          \path [#1]
          (\tikzinputsegmentfirst) -- (\tikzinputsegmentlast);
        },
      },
    },
    % style to add an arrow in the middle of a path
    mid arrow/.style={postaction={decorate,decoration={
          markings,
          mark=at position .6 with {\arrow[#1]{stealth}}
        }}},
    small mid arrow/.style={postaction={decorate,decoration={
          markings,
          mark=at position .5 with {\arrow[#1]{stealth'}}
        }}},
    boson_wdir/.style={draw=black, dashed, postaction={decorate}, decoration={markings,mark=at position .5 with {\arrow[draw=black]{>}}}},
    boson/.style={draw=black, dashed},
    photon/.style={decorate,decoration={snake},draw=black},
    gluon/.style={decorate,decoration={gluon_decoration, aspect=1.5, amplitude=1.5pt, segment length=8pt}},
    gluon_up/.style={decorate,decoration={gluon_decoration, mirror, aspect=1.5, amplitude=1.5pt, segment length=8pt}},
    fermion/.style={mid arrow}
  }

  \coordinate (feynman-glu-in) at (0,0);
  \coordinate (feynman-vtx) at ($(feynman-glu-in) + (1.5em, 0em)$);
  \coordinate (feynman-q-out) at ($(feynman-vtx) + (1.5em, 0.5em)$);
  \coordinate (feynman-qbar-out) at ($(feynman-vtx) + (1.5em, -0.5em)$);

  \coordinate (connector-lhs) at ($(feynman-vtx) + (1.6em, 0em)$);
  \coordinate (connector-rhs) at ($(connector-lhs) + (1em, 0em)$);

  \coordinate (cflow-glu-in) at ($(connector-rhs) + (0.25em, 0em)$);
  \coordinate (cflow-vtx) at ($(cflow-glu-in) + (1.4em, 0em)$);
  
  \draw[gluon] (feynman-glu-in) -- (feynman-vtx);
  \draw[postaction={on each segment={fermion}}, line join=bevel] (feynman-qbar-out) -- (feynman-vtx) -- (feynman-q-out);

  \draw[-implies,double equal sign distance] (connector-lhs) -- (connector-rhs);

  \draw[Red, postaction={on each segment={small mid arrow={Red, scale=0.6}}}]
  ($(cflow-glu-in) + (0em, +0.1em)$) -- ($(cflow-vtx) + (0em, +0.1em)$) -- ++(+20:1.05em);

  \draw[Green, postaction={on each segment={small mid arrow={Green, scale=0.6, rotate=180}}}]
  ($(cflow-glu-in) + (0em, -0.1em)$) -- ($(cflow-vtx) + (0em, -0.1em)$) -- ++(-20:1.05em);
\end{tikzpicture}}
    % \rule{\columnwidth}{3pt}
  \end{subfigure}
  \hfill
  \begin{subfigure}{0.4\textwidth}
    \centering
    \scalebox{1.6}{\begin{tikzpicture}[thick]
  \tikzset{
    % style to apply some styles to each segment of a path
    on each segment/.style={
      decorate,
      decoration={
        show path construction,
        moveto code={},
        lineto code={
          \path [#1]
          (\tikzinputsegmentfirst) -- (\tikzinputsegmentlast);
        },
        curveto code={
          \path [#1] (\tikzinputsegmentfirst)
          .. controls
          (\tikzinputsegmentsupporta) and (\tikzinputsegmentsupportb)
          ..
          (\tikzinputsegmentlast);
        },
        closepath code={
          \path [#1]
          (\tikzinputsegmentfirst) -- (\tikzinputsegmentlast);
        },
      },
    },
    % style to add an arrow in the middle of a path
    mid arrow/.style={postaction={decorate,decoration={
          markings,
          mark=at position .6 with {\arrow[#1]{stealth}}
        }}},
    small mid arrow/.style={postaction={decorate,decoration={
          markings,
          mark=at position .5 with {\arrow[#1]{stealth'}}
        }}},
    boson_wdir/.style={draw=black, dashed, postaction={decorate}, decoration={markings,mark=at position .5 with {\arrow[draw=black]{>}}}},
    boson/.style={draw=black, dashed},
    photon/.style={decorate,decoration={snake},draw=black},
    gluon/.style={decorate,decoration={gluon_decoration, aspect=1.5, amplitude=1.5pt, segment length=8pt}},
    gluon_up/.style={decorate,decoration={gluon_decoration, mirror, aspect=1.5, amplitude=1.5pt, segment length=8pt}},
    fermion/.style={mid arrow}
  }

  \coordinate (feynman-glu-in) at (0,0);
  \coordinate (feynman-vtx) at ($(feynman-glu-in) + (1.5em, 0em)$);
  \coordinate (feynman-glu1-out) at ($(feynman-vtx) + (1.5em, 0.65em)$);
  \coordinate (feynman-glu2-out) at ($(feynman-vtx) + (1.5em, -0.65em)$);

  \coordinate (connector-lhs) at ($(feynman-vtx) + (1.6em, 0em)$);
  \coordinate (connector-rhs) at ($(connector-lhs) + (1em, 0em)$);

  \coordinate (cflow-glu-in) at ($(connector-rhs) + (0.25em, 0em)$);
  \coordinate (cflow-vtx) at ($(cflow-glu-in) + (1.1em, 0em)$);
  \coordinate (cflow-vtx-out) at ($(cflow-vtx) + (0.3em, 0em)$);

  \coordinate (connector2) at ($(cflow-vtx-out) + (1.2em, 0em)$);

  \coordinate (cflow2-glu-in) at ($(connector2) + (0.25em, 0em)$);
  \coordinate (cflow2-vtx) at ($(cflow2-glu-in) + (1.1em, 0em)$);
  \coordinate (cflow2-vtx-out) at ($(cflow2-vtx) + (0.4em, 0em)$);
  
  \draw[gluon] (feynman-glu-in) -- (feynman-vtx);
  \draw[gluon] (feynman-vtx) -- (feynman-glu1-out);
  \draw[gluon_up] (feynman-vtx) -- (feynman-glu2-out);

  \draw[-implies,double equal sign distance] (connector-lhs) -- (connector-rhs);

  \draw[Red, postaction={on each segment={small mid arrow={Red, scale=0.6}}}]
  ($(cflow-glu-in) + (0em, +0.1em)$) -- ($(cflow-vtx) + (0em, +0.1em)$) -- ++(+25:1.22em);

  \draw[Green, postaction={on each segment={small mid arrow={Green, scale=0.6, rotate=180}}}]
  ($(cflow-glu-in) + (0em, -0.1em)$) -- ($(cflow-vtx) + (0em, -0.1em)$) -- ++(-25:1.22em);

  \draw[Cyan, postaction={on each segment={small mid arrow={Cyan, scale=0.6}}}]
  ($(cflow-vtx-out) + (+25:1.05em)$) -- (cflow-vtx-out) -- ++(-25:1.05em);

  \node (connector2-txt) at ($(connector2) + (0em, -0.2em)$) {,};
  
  % \draw[Red, postaction={on each segment={small mid arrow={Red, scale=0.6}}}]
  % ($(cflow2-glu-in) + (0em, +0.1em)$) -- ($(cflow2-vtx) + (0.25em, +0.1em)$) -- ++(-30:1.3em);

  % \draw[Green, postaction={on each segment={small mid arrow={Green, scale=0.6, rotate=180}}}]
  % ($(cflow2-glu-in) + (0em, -0.1em)$) -- ($(cflow2-vtx) + (0.25em, -0.1em)$) -- ++(+30:1.3em);

  % \draw[Cyan, postaction={on each segment={small mid arrow={Cyan, scale=0.6}}}]
  % ($(cflow2-vtx-out) - (0.2em, 0em) + (+30:1.5em)$) -- ($(cflow2-vtx-out) - (0.2em, 0em)$) -- ++(-30:1.5em);

  \draw[Red] ($(cflow2-glu-in) + (0em, +0.1em)$) -- ($(cflow2-vtx) + (-0.3em, +0.1em)$) -- ++(-25:1.5em);
  \draw[Red, small mid arrow={Red, scale=0.6}] ($(cflow2-glu-in) + (0em, +0.1em)$) -- ($(cflow2-vtx) + (-0.3em, +0.1em)$);
  \draw[Red, postaction={decorate,decoration={markings, mark=at position .7 with {\arrow[Red, scale=0.6]{stealth'}}}}] ($(cflow2-vtx) + (-0.3em, +0.1em)$) -- ++(-25:1.5em);

  \draw[Green] ($(cflow2-glu-in) + (0em, -0.1em)$) -- ($(cflow2-vtx) + (-0.3em, -0.1em)$) -- ++(+25:1.5em);
  \draw[Green, postaction={on each segment={small mid arrow={Green, scale=0.6, rotate=180}}}] ($(cflow2-glu-in) + (0em, -0.1em)$) -- ($(cflow2-vtx) + (-0.3em, -0.1em)$);
  \draw[Green, postaction={decorate,decoration={markings, mark=at position .7 with {\arrow[Green, scale=0.6, rotate=180]{stealth'}}}}] ($(cflow2-vtx) + (-0.3em, -0.1em)$) -- ++(+25:1.5em);

  \draw[Cyan, postaction={on each segment={small mid arrow={Cyan, scale=0.6, rotate=180}}}]
  ($(cflow2-vtx-out) + (+25:0.9em)$) -- (cflow2-vtx-out) -- ++(-25:0.9em);

\end{tikzpicture}}
    % \rule{\columnwidth}{3pt}
  \end{subfigure}
  \caption{QCD colour propagation rules for elementary quark--gluon vertices.
    Black lines denote Feynman-diagram style vertices, coloured lines show QCD
    colour connection lines.}
  \label{fig:colour-propagation-rules}
\end{figure}

In high-energy hadron collisions, such as those produced at the Large Hadron
Collider (LHC)~\cite{Evans:2008zzb} at CERN, quarks and gluons are produced
abundantly. However, due to the confining nature of quantum chromodynamics
(QCD), the direct measurement of the interactions that occur between these
particles is impossible and only colour-neutral hadrons can be measured. % The
% strength and direction of the strong force depends on the colour charge of the
% particles involved.
To a good approximation, the radiation pattern in QCD can be
described through a colour–connection picture, which consists of colour strings
connecting quarks and gluons of one colour to quarks and gluons of the
corresponding anti–colour. % An important question is whether there is evidence of
% these colour connections (colour flow) in the observable objects: colour–neutral
% hadrons and the jets they form.
% The quarks and gluons of the hard-scatter
% interaction carry colour charge, which is conserved for every subsequent
% interaction or decay.
Figure~\ref{fig:colour-propagation-rules} illustrates the
colour connections for the relevant elementary QCD vertices.

In the decay chain of a hard-scatter event, the colour charge ``flows'' from the
initial state towards stable particles whilst following the rules illustrated in
Figure~\ref{fig:colour-propagation-rules}. As colour charge is conserved,
connections exist between initial particles and the stable colour-neutral
hadrons.

In practice, high-energy quarks and gluons are measured as jets, which are
bunches of collimated hadrons that form in the evolution of the coloured initial
particles. The colour connections between high-energy particles affect the
structure of the emitted radiation and therefore also the structure of the resulting
jets. For example, soft gluon radiation is suppressed in some regions of phase
space compared to others. Specifically, due to colour coherence effects, QCD
predicts an increase of radiation where a colour connection is present compared to a region
of phase space where no such connection exists, see Ref.~\cite{Ellis:318585}.
Smaller effects on the event topology and measured quantities are expected from
colour reconnection in the hadronisation process.

Providing evidence for the existence of the connections between particles~---~the \textit{colour flow}~---~is important for the validation of
phenomenological descriptions. Using the energy-weighted distributions of particles within and between jets has
been a long-standing tool for investigating colour flow, with early measurements
at PETRA~\cite{Bartel1983} and LEP~\cite{ACHARD2003202,Abdallah:2006uq}.
Later, a precursor of the jet pull was
studied using the abundant jet production at the Tevatron~\cite{1999145}. Recently, the colour flow was measured by ATLAS in \ttbar events
at the LHC at a centre-of-mass energy of
$\sqrt{s}=\SI{8}{\TeV}$~\cite{TOPQ-2014-09} using the jet-pull angle.

\begin{figure}[!htb]
  \centering
  % \selectcolormodel{gray}
  \scalebox{1.5}{\begin{tikzpicture}[thick]
  \tikzset{
    % style to apply some styles to each segment of a path
    on each segment/.style={
      decorate,
      decoration={
        show path construction,
        moveto code={},
        lineto code={
          \path [#1]
          (\tikzinputsegmentfirst) -- (\tikzinputsegmentlast);
        },
        curveto code={
          \path [#1] (\tikzinputsegmentfirst)
          .. controls
          (\tikzinputsegmentsupporta) and (\tikzinputsegmentsupportb)
          ..
          (\tikzinputsegmentlast);
        },
        closepath code={
          \path [#1]
          (\tikzinputsegmentfirst) -- (\tikzinputsegmentlast);
        },
      },
    },
    % style to add an arrow in the middle of a path
    mid arrow/.style={postaction={decorate,decoration={
          markings,
          mark=at position .6 with {\arrow[#1]{stealth}}
        }}},
    small mid arrow/.style={postaction={decorate,decoration={
          markings,
          mark=at position .5 with {\arrow[#1]{stealth'}}
        }}},
    boson_wdir/.style={draw=black, dashed, postaction={decorate}, decoration={markings,mark=at position .5 with {\arrow[draw=black]{>}}}},
    boson/.style={draw=black, dashed},
    photon/.style={decorate,decoration={snake},draw=black},
    gluon/.style={decorate,decoration={gluon_decoration, aspect=1.5, amplitude=2.5pt, segment length=8pt}},
    gluon_up/.style={decorate,decoration={gluon_decoration, mirror, aspect=1.5, amplitude=1.5pt, segment length=8pt}},
    fermion/.style={mid arrow}
  }

  \coordinate (feynman-glu-in) at (0,0);
  \coordinate (feynman-glu-ttbar) at ($(feynman-glu-in) + (4em, 0em)$);

  \coordinate (feynman-t-decay) at ($(feynman-glu-ttbar) + (3.3em, 1.75em)$);
  \coordinate (feynman-tbar-decay) at ($(feynman-glu-ttbar) + (3.3em, -1.75em)$);

  \coordinate (feynman-b) at ($(feynman-t-decay) + (5em, 1em)$);
  \coordinate (feynman-bbar) at ($(feynman-tbar-decay) + (5em, -1em)$);

  \coordinate (feynman-Wp-decay) at ($(feynman-t-decay) + (2.5em, -1em)$);
  \coordinate (feynman-Wm-decay) at ($(feynman-tbar-decay) + (2.5em, 1em)$);

  \coordinate (feynman-l+) at ($(feynman-Wp-decay) + (2.5em, +1.0em)$);
  \coordinate (feynman-nu) at ($(feynman-Wp-decay) + (2.5em, -0.3em)$);

  \coordinate (feynman-q) at ($(feynman-Wm-decay) + (2.5em, +0.3em)$);
  \coordinate (feynman-qbar) at ($(feynman-Wm-decay) + (2.5em, -1.0em)$);

  \draw[gluon] (feynman-glu-in) -- (feynman-glu-ttbar);
  \draw (feynman-glu-ttbar) to node[midway, fill=white, inner sep=1pt] {\scriptsize $t$} (feynman-t-decay);
  \draw (feynman-glu-ttbar) to node[midway, fill=white, inner sep=1pt] {\scriptsize $\bar{t}$} (feynman-tbar-decay);

  \draw (feynman-t-decay) to node[midway, fill=white, inner sep=1pt] {\scriptsize $W^+$} (feynman-Wp-decay);
  \draw (feynman-tbar-decay) to node[midway, fill=white, inner sep=1pt] {\scriptsize $W^-$} (feynman-Wm-decay);

  \draw (feynman-t-decay) to (feynman-b);
  \draw (feynman-tbar-decay) to (feynman-bbar);

  \draw (feynman-Wp-decay) to (feynman-l+);
  \draw (feynman-Wp-decay) to (feynman-nu);

  \draw (feynman-Wm-decay) to (feynman-q);
  \draw (feynman-Wm-decay) to (feynman-qbar);

  \node[anchor=west] at ($(feynman-b) + (0em, 0em)$) {\scriptsize $b$};
  \node[anchor=west] at ($(feynman-bbar) + (0em, 0em)$) {\scriptsize $\bar{b}$};

  \node[anchor=west] at ($(feynman-l+) + (0em, 0em)$) {\scriptsize $\ell^+$};
  \node[anchor=west] at ($(feynman-nu) + (0em, 0em)$) {\scriptsize $\nu_{\ell}$};

  \node[anchor=west] at ($(feynman-q) + (0em, 0em)$) {\scriptsize $q$};
  \node[anchor=west] at ($(feynman-qbar) + (0em, 0em)$) {\scriptsize $\bar{q}^{\prime}$};

  % Draw colour lines
  \draw[Red, ultra thick, rounded corners] ($(feynman-glu-in) + (0em, 0.5em)$) to
  ($(feynman-glu-ttbar) + (0em, 0.5em)$) to
  ($(feynman-t-decay) + (0em, 0.5em)$) to
  ($(feynman-b) + (0em, 0.5em)$);

  \draw[Green, ultra thick, rounded corners] ($(feynman-glu-in) + (0em, -0.5em)$) to
  ($(feynman-glu-ttbar) + (0em, -0.5em)$) to
  ($(feynman-tbar-decay) + (0em, -0.5em)$) to
  ($(feynman-bbar) + (0em, -0.5em)$);

  % Helper line
  \path[name path=cline-helper-q] ($(feynman-Wm-decay) + (0em, -0.2em)$) -- ($(feynman-q) + (0em, -0.2em)$);
  \path[name path=cline-helper-qbar] ($(feynman-Wm-decay) + (0em, 0.2em)$) -- ($(feynman-qbar) + (0em, +0.2em)$);
  \path [name intersections={of=cline-helper-q and cline-helper-qbar, by=cvtx-q-qbar}];

  \draw[Cyan, ultra thick, rounded corners] ($(feynman-q) + (0em, -0.2em)$) to
  (cvtx-q-qbar) to
  ($(feynman-qbar) + (0em, +0.2em)$);

  % Draw vertices
  \node[draw, shape=circle, fill=black, minimum size=0.2em, inner sep=0pt] at (feynman-glu-ttbar) {};
  \node[draw, shape=circle, fill=black, minimum size=0.2em, inner sep=0pt] at (feynman-t-decay) {};
  \node[draw, shape=circle, fill=black, minimum size=0.2em, inner sep=0pt] at (feynman-tbar-decay) {};
  \node[draw, shape=circle, fill=black, minimum size=0.2em, inner sep=0pt] at (feynman-Wp-decay) {};
  \node[draw, shape=circle, fill=black, minimum size=0.2em, inner sep=0pt] at (feynman-Wm-decay) {};
\end{tikzpicture}}
  \caption{Illustration of a semileptonic \ttbar event with typical colour
    connections (thick coloured lines).}
  \label{fig:illustration-ttbar}
\end{figure}
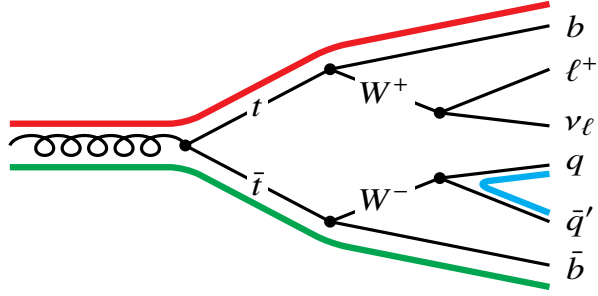

Figure~\ref{fig:illustration-ttbar} illustrates the production of a \ttbar pair
and its subsequent decay into a single-lepton final state as produced at the LHC
with colour connections superimposed. In the hard-scatter event, four
colour-charged final states can be identified: the two $b$-quarks produced
directly by the decay of the top-quarks and the two quarks produced by the
hadronically decaying $W$ boson. As the $W$ boson does not carry colour charge,
its daughters must share a colour connection. The two $b$-quarks from the
top-quark decays carry the colour charge of their respective top-quark parent,
and are thus not expected to share a colour connection.

Despite the long-standing history of measurements of
the potential effects of  colour connections,
they remain a poorly constrained effect of QCD and require further
experimental input. Furthermore, it may be possible to use
the extracted colour information to distinguish between event topologies with
a different colour structure. In the case of jets, such colour information would complement the
kinematic properties, and might enable the identification of otherwise irreducible
backgrounds, or facilitate the correct assignment of jets to a particular physical
process. For example, a colour-flow observable could be used to resolve the
ambiguity in assigning $b$-jets to the Higgs boson decay in $t\bar{t}H(\to
b\bar{b})$ events.

An observable predicted to encode colour information about a jet
is the jet-pull vector $\vec{\mathcal{P}}$~\cite{Gallicchio:2010sw}, a
\pt-weighted radial moment of the jet. For a given jet $j$ with
transverse momentum $\pt^j$, the observable is defined as

\begin{align}
  \vec{\mathcal{P}}\left( j \right) = \sum_{i \in j} \frac{\left| \vec{\Delta r}_{i} \right| \cdot \pt^i}{\pt^j} \vec{\Delta r}_i \, ,
  \label{eq:jet-pull}
\end{align}
where the summation runs over the constituents of $j$ that have transverse
momentum $\pt^i$ and are located at $\vec{\Delta r}_i = \left( \Delta y_i, \Delta
  \phi_i \right)$, which is the offset of the constituent from the jet axis
$(y_j, \phi_j)$ in rapidity--azimuth ($y$--$\phi$) space.\footnote{ATLAS
  uses a right-handed coordinate system with its origin at the nominal
  interaction point (IP) in the centre of the detector and the $z$-axis along
  the beam pipe. The $x$-axis points from the IP to the centre of the LHC ring,
  and the $y$-axis points upward. Cylindrical coordinates $(r,\phi)$ are used in
  the transverse plane, $\phi$ being the azimuthal angle around the $z$-axis.
  The rapidity, which is used in the jet-pull vector calculation, is defined as
  $y = \frac12 \ln \frac{E + p_z}{E - p_z}$ using an object's energy $E$ and
  momentum $p_z$ along the $z$-axis. A related quantity is the
  pseudorapidity, which is defined in terms of the polar angle $\theta$ as
  $\eta=-\ln\tan(\theta/2)$. Using these coordinates, the radial distance
  $\Delta R$ between two objects is thus defined as $\Delta R = \sqrt{(\Delta\eta)^2
    + (\Delta\phi)^2}$ where $\Delta\eta$ and $\Delta\phi$ are the differences in
  pseudorapidity and azimuthal angle between the two objects, respectively.}
Examples of constituents that could be used in Eq.~\eqref{eq:jet-pull} include
calorimeter energy clusters, inner-detector tracks, and simulated stable particles.

\begin{figure}[!htb]
  \centering
  \includegraphics[width=.45\textwidth]{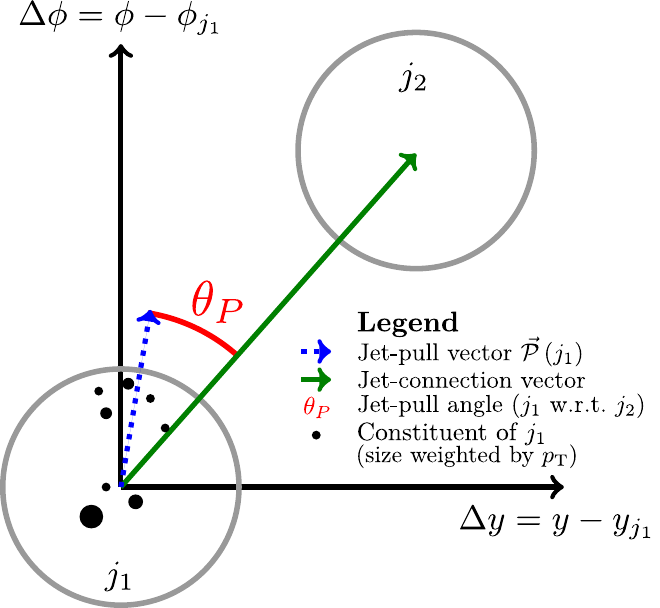}
  \caption{Illustration of jet-pull observables for a dijet system. For a jet
    $j_1$ the jet-pull vector is calculated using an appropriate
    set of constituents (tracks, calorimeter energy clusters, simulated particles, \ldots).
    The variable of particular sensitivity to the colour structure of $j_1$ with
    respect to $j_2$ is the jet-pull angle $\theta_P$ which is the angle between the
    pull vector for $j_1$ and the vector connecting $j_1$ to another jet $j_2$
    in localised $y$--$\phi$ space (the ``jet connection vector'').}
  \label{fig:jet-pull-schema}
\end{figure}

Given two jets, $j_1$ and $j_2$, the jet-pull vector can be used to construct
the jet-pull angle $\theta_{\mathcal{P}}\left( j_1, j_2 \right)$. This is
defined as the angle between the jet-pull vector $\vec{\mathcal{P}}\left( j_1
\right)$ and the vector connecting $j_1$ to $j_2$ in rapidity--azimuth space,
$\left( y_{j_2} - y_{j_1}, \phi_{j_2} - \phi_{j_1} \right)$, which is called
``jet connection vector''. Figure~\ref{fig:jet-pull-schema} illustrates the
jet-pull vector and angle for an idealised dijet system. As the jet-pull angle
is symmetric around zero and takes values ranging from $-\pi$ to $\pi$, it is
convenient to consider the normalised absolute pull angle $\left|
\theta_{\mathcal{P}} \right| / \pi$ as the observable. The measurement presented
here is performed using this normalisation.

The jet-pull angle is particularly suited for studying the colour
structure of an object decaying to a dijet system,
as the inputs into the calculation are well-defined theoretically and the
observable is expected to be sensitive to the presence or absence of a colour
connection. For two colour-connected jets, $j_1$ and $j_2$, it is expected that
$\vec{\mathcal{P}}\left( j_1 \right)$ and $\vec{\mathcal{P}}\left( j_2 \right)$
are aligned with the jet connection vector, i.e.\ $\theta_{\mathcal{P}} \sim 0$.
For two jets without any particular colour connection, the jet-pull vector and
the connection vector are not expected to be aligned and thus
$\theta_{\mathcal{P}}$ is expected to be distributed uniformly.

In this paper, the normalised jet-pull angle is measured for two different
systems of dijets in \ttbar events using $36.1\,\text{fb}^{-1}$ of $pp$
collision data recorded by the ATLAS detector at $\sqrt{s} = 13 \, \text{TeV}$.
The first targets the jets originating from the hadronic decay of a $W$ boson
and thus from a colour singlet, while the second targets the two $b$-jets from
the top decays, which are not expected to be colour connected. The
magnitude of the jet-pull vector is also measured. The results are presented as
normalised distributions corrected for detector effects.

In Section~\ref{sec:atlas-detector}, the ATLAS detector is introduced.
Section~\ref{sec:data-simulation} discusses the data and simulation samples
used by this
analysis. The reconstruction procedures and
event selection are presented in Section~\ref{sec:event-reco-selection}.
In Section~\ref{sec:observables} the analysis observables are
introduced and discussed in detail. Section~\ref{sec:unfolding} introduces the
phase space of the particle-level measurement and the unfolding procedure used
to correct the observed data for detector effects.
In Section~\ref{sec:systematics} the relevant uncertainties and
the methodology used to assess them are discussed. Finally,
Section~\ref{sec:result} presents the results, followed by a conclusion in
Section~\ref{sec:conclusion}.

% -------------------------------------------------------------------------------
\section{The ATLAS detector}
\label{sec:atlas-detector}
% -------------------------------------------------------------------------------

The ATLAS detector~\cite{PERF-2007-01} is a multi-purpose detector with a near
$4\pi$ coverage in solid angle. It uses a system of tracking detectors, which
enclose the interaction point, to provide highly resolved spatial
measurements of charged particles in the range $\left| \eta \right| <
2.5$. These tracking detectors, collectively called the inner detector, are immersed
in a \SI{2}{\tesla} magnetic field enabling reconstruction of the track
momentum. During the Long Shutdown 1, a new innermost layer of
the pixel detector was inserted into the detector, the insertable B-layer
(IBL)~\cite{ATLAS-TDR-19,CERN-LHCC-2012-009}. Two calorimeter subsystems
enclose the inner detector allowing complementary calorimetric measurements of
both the charged and neutral particles. Behind the calorimeters a system of 
muon chambers provides muon identification, triggering, and
(additional) tracking. The muon system is immersed in a magnetic field provided
by three toroid magnets. A more complete description of the ATLAS detector can be found
elsewhere~\cite{PERF-2007-01}.

Data are selected for read-out and further processing by a two-stage
trigger~\cite{TRIG-2016-01} that uses coarse detector information in a hardware-based
first stage followed by a
software-based second trigger stage, which has access to the full detector
granularity. This reduces the raw rate of \SI{40}{\mega\Hz}
from the LHC $pp$ collisions to about \SI{75}{\kilo\Hz} after the
first stage and \SI{1}{\kilo\Hz} after the second stage.

% -------------------------------------------------------------------------------
\section{Data sample and simulation}
\label{sec:data-simulation}
% -------------------------------------------------------------------------------

The data used by this analysis were collected in 2015 and 2016 during $pp$
runs provided by the LHC at a centre-of-mass energy of
$\sqrt{s}=\SI{13}{\TeV}$. Stable beams and fully operational subdetectors are
required. After data quality requirements, the data correspond to an integrated
luminosity of $\mathcal{L}_{\text{Int}} = \SI{36.1}{\per\femto\barn}$.

Monte Carlo (MC) samples are used to evaluate the contribution of background
processes to the selected event sample,
evaluate how the detector response affects
the analysis observables and for comparisons with the measured data.
A variety of configurations are investigated for
different purposes. Table~\ref{tab:mc-samples} summarises the samples used by
the analysis.

The $t\bar{t}$ sample in the first row of the table (the
\textit{``nominal''} sample) is used to evaluate how well the data agrees with MC simulation,
predict the number of signal events, and obtain the nominal detector response
description. This sample was generated using the
\POWHEGBOX~\texttt{v2}~\cite{powheg1,powheg2,powheg3} event generator with
the \texttt{NNPDF} parton distribution functions (PDF)~\cite{NNPDF}. The
top-quark mass, $m_t$, was set to $\SI{172.5}{\GeV}$ and the value of the
$h_{\mathrm{damp}}$ parameter, which controls the \pt of the first emission
beyond the Born configuration in \POWHEG, was set to $1.5~m_t$. The main effect
of $h_{\mathrm{damp}}$ is to regulate the high-\pt emission against which the \ttbar system
recoils. \PYTHIAV{8}~\cite{pythia8} with the \texttt{NNPDF}~\cite{Ball:2012cx}
PDF set and the \texttt{A14}~\cite{ATL-PHYS-PUB-2014-021}
tune\footnote{\label{tune_footnote} The term \textit{tune} refers to a specific
setting of configurable parameters of the MC generator describing non-perturbative QCD effects. A tune variation can be
used to assess the effect of the modelling of non-perturbative effects on an analysis.} was used to simulate the parton shower, hadronisation and
underlying event.

\begin{table}[!bth]
  \centering
  \scriptsize 
  \setlength{\tabcolsep}{4pt}
  \begin{tabularx}{\textwidth}{llllll}
    \toprule
    \textbf{Process} & \textbf{Generator} & \textbf{Type} & \textbf{Version} & \textbf{PDF} & \textbf{Tune}\footref{tune_footnote} \\
    \midrule
    $t\bar{t}$ & \POWHEGBOX~\texttt{v2}~\cite{powheg1,powheg2,powheg3} & NLO ME & \texttt{r3026} & \texttt{NNPDF 3.0}~\cite{NNPDF} & -- \\
    & +\PYTHIAV{8}~\cite{pythia8} & +LO PS & \texttt{v8.186} & \texttt{NNPDF 2.3}~\cite{Ball:2012cx} & \texttt{A14}\,/\,\texttt{A14.v1}$^{\dagger}$\,/\,\texttt{A14.v3c}$^{\dagger}$~\cite{ATL-PHYS-PUB-2014-021} \\
    
    &&&&& \\[0.4em]
    $t\bar{t}^{\dagger}$ & \POWHEGBOX~\texttt{v2} & NLO ME & \texttt{r3026} & \texttt{NNPDF 3.0} & -- \\
    & +\HERWIGV{7}~\cite{Bellm:2015jjp} & +LO PS & \texttt{v7.0.1.a} & \texttt{MMHT 2014}~\cite{MMHT} & \texttt{H7UE} \\
    &&&&& \\[-0.4em]
    $t\bar{t}^{\dagger}$ & \MGMCatNLO~\cite{Alwall:2011uj} & NLO ME & \texttt{v2.3.3.p1} & \texttt{NNPDF 3.0} & -- \\
    & +\PYTHIAV{8} & +LO PS & \texttt{v8.112} & \texttt{NNPDF 2.3} & \texttt{A14} \\
    &&&&& \\[0.4em]
    $t\bar{t}^{\star}$ & \POWHEGBOX~\texttt{v2} & NLO ME & \texttt{r2819} & \texttt{CT10}~\cite{CT10} & -- \\
    & +\PYTHIAV{6}~\cite{pythia6} & +LO PS & \texttt{v6.428} & \texttt{CTEQ6L1}~\cite{Pumplin:2002vw} & \textsc{Perugia 2012}~\cite{Skands:2010ak} \\
    &&&&& \\[-0.4em]
    $t\bar{t}^{\star}$ &
    \SHERPA~\cite{Gleisberg:2008ta,Schumann:2007mg,sherpanlo} & \makecell[l]{LO/NLO \\ multileg ME+PS} & \texttt{v2.2.1} & \texttt{NNPDF 3.0
      NNLO} & -- \\
    
    \midrule[0.5pt]
    
    Single top & \POWHEGBOX~\texttt{v1} & NLO ME & \texttt{r2819} & \texttt{CT10} & -- \\
    & +\PYTHIAV{6} & +LO PS & \texttt{v6.425} & \texttt{CTEQ6L1} & \textsc{Perugia 2012} \\

    &&&&& \\[0.4em]

    $WW,\,WZ,\,ZZ$ & \SHERPA & \makecell[l]{LO/NLO \\ multileg ME+PS} & \texttt{v2.1.1} & \texttt{CT10} & Default \\

    &&&&& \\[0.4em]

    $W/Z+\text{jets}$ & \SHERPA & \makecell[l]{LO/NLO \\ multileg ME+PS} & \texttt{v2.2.1} & \texttt{NNPDF 3.0} & Default \\
    &&&&& \\[0.4em]

    $t\bar{t}W/Z$ & \MGMCatNLO & NLO ME & \texttt{v2.3.3} & \texttt{NNPDF 3.0} & -- \\
    & +\PYTHIAV{8}~\cite{Sjostrand:2014zea} & +LO PS & \texttt{v8.210} & \texttt{NNPDF 2.3} & \texttt{A14} \\
    &&&&& \\[0.4em]
    $t\bar{t}H$ & \MGMCatNLO & NLO ME & \texttt{v2.2.3.p4} & \texttt{NNPDF 3.0} & -- \\
    & +\PYTHIAV{8} & +LO PS & \texttt{v8.210} & \texttt{NNPDF 2.3} & \texttt{A14} \\
    \bottomrule
  \end{tabularx}
  \caption{Monte Carlo samples used for this analysis. The first part of the
    table shows samples generated for the signal process, the second those for
    processes considered to be a background. Samples / tunes marked with
    $\dagger$ refer to alternative signal MC samples used to evaluate signal
    modelling uncertainties, those marked with $\star$ are used for comparison
    to the measurement result. The following abbreviations are used:
    \textit{ME} -- matrix element, \textit{PS} -- parton shower, \textit{LO} --
    leading-order calculation in QCD, \textit{NLO} -- next-to-leading-order
    calculation in QCD, \textit{PDF} -- parton distribution function.
  }
  \label{tab:mc-samples}
\end{table}

To evaluate the impact of systematic uncertainties coming from signal modelling
on the measurements, a variety of alternative signal MC samples are used. These
samples or tunes are marked with a $\dagger$ in Table~\ref{tab:mc-samples}. To assess the
impact of increased or reduced radiation, samples were generated using the
\texttt{A14.v3c} up and down tune variations. Additionally, in the
\texttt{A14.v3c} up (down) variation sample the renormalisation and
factorisation scales were scaled by a factor of 0.5 (2) relative to the
nominal sample and the value of $h_{\mathrm{damp}}$ was set to $3 m_t$ ($1.5
m_t$)~\cite{ATL-PHYS-PUB-2016-020}. Similarly, to assess the impact of colour reconnection, two samples generated
with the \texttt{A14.v1} tune variations are used. These modify simulation
parameters which configure the strong coupling of multi-parton interactions and
the strength of the colour-reconnection mechanism~\cite{ATL-PHYS-PUB-2014-021}. Two alternative MC programs are used in order to
estimate the impact of the choice of hard-scatter generator and hadronisation
algorithm: for each of these samples one of the two components is replaced by an
alternative choice. The alternative choices are
\MGMCatNLO~(MG5\_aMC)~\cite{Alwall:2011uj} for the hard-scatter generator and
\HERWIGV{7}~\cite{Bellm:2015jjp} for the hadronisation algorithm.

Two additional simulation set-ups are used to obtain \ttbar predictions,
both of which are marked with a $\star$ in Table~\ref{tab:mc-samples}: one
sample uses \POWHEGBOX~\texttt{v2}, with $h_{\mathrm{damp}}$ set to
the top-quark mass, interfaced to \PYTHIAV{6} for the hadronisation. The
second set-up uses the \SHERPA~\cite{Gleisberg:2008ta,Schumann:2007mg,sherpanlo}
MC program to construct predictions from theoretical calculation.

Signal MC simulation is normalised to a cross-section of $832^{+46}_{-51} \,
\si{\pico\barn}$, where the uncertainties reflect
the effect of scale, PDF, and $\alpha_s$ variations as well as the top-quark
mass uncertainty. This is calculated with the \texttt{Top++ 2.0} program~\cite{CZAKON20142930} to
next-to-next-to-leading order in perturbative QCD, including resummation of
next-to-next-to-leading-logarithm soft-gluon terms, assuming a top-quark mass of
$\SI{172.5}{\GeV}$~\cite{Cacciari:2011hy,Beneke:2011mq,Baernreuther:2012ws,Czakon:2012zr,Czakon:2012pz,Czakon:2013goa}.
Normalised signal MC simulation is only used to compare the observed data to the
prediction.

Contributions from processes considered to be a background to the analysis are
in most cases modelled using simulation samples. These samples are shown in the
second part of Table~\ref{tab:mc-samples}. All background MC samples are
normalised to their theoretical cross-sections evaluated to at least
next-to-leading order (NLO) precision in
QCD~\cite{PhysRevLett.103.082001,PhysRevD.81.054028,PhysRevD.82.054018,PhysRevD.83.091503,PhysRevD.60.113006,Campbell:2011bn,Alwall:2014hca,deFlorian:2016spz,ATL-PHYS-PUB-2016-003,deFlorian:2016spz}.

Multiple overlaid $pp$ collisions, which are causing so called pile-up, were simulated with the soft QCD
processes of \PYTHIAV{8.186}~\cite{pythia8} using the
\texttt{A2}~\cite{ATL-PHYS-PUB-2012-003} tune and the \texttt{MSTW2008LO} PDF
set~\cite{Martin:2009iq}. A reweighting procedure was applied on an
event-by-event basis to the simulation samples to reflect the distribution of
the average number of $pp$ interactions per event observed in data.

Events generated by the MC programs are further processed using the ATLAS
detector and trigger simulation~\cite{SOFT-2010-01} which uses
{\GEANT}4~\cite{AGOSTINELLI2003250} to simulate the interactions between
particles and the detector material. The samples used to evaluate the detector
response and estimate the background contributions were processed using the full
ATLAS simulation~\cite{SOFT-2010-01}. Alternative signal MC samples, which are
used to evaluate signal modelling uncertainties, were processed using
\texttt{Atlfast II}~\cite{Richter-Was:683751}. This detector simulation differs
from the full ATLAS detector simulation by using a faster method to model energy
depositions in the calorimeter, while leaving the simulation of the remainder of
the detector unchanged.

In order to evaluate the sensitivity of the analysis observables to colour flow
and to be able to assess the colour-model dependence of the analysis methods, a
dedicated MC sample with a simulated exotic colour-flow model is used; this is
labelled as \textit{``(colour) flipped''}. In this sample, the colour-singlet $W$
boson in ordinary signal events is replaced \textit{ad hoc} by a colour octet.
To create this sample, hard-scatter signal events were generated using
\POWHEGBOX~\texttt{v2} with the same settings as the nominal \ttbar sample and
stored in the LHE format~\cite{Alwall:2006yp}. The colour strings were then
flipped in such a way that, among the decay products obtained from the hadronic
decay of the $W$~boson, one of them is connected to the incoming top quark while
the other one is connected to the outgoing $b$-quark. \PYTHIAV{8} was then used
to perform the showering and hadronisation in the modified hard-scatter event
using the same procedure as in the nominal \ttbar sample.

% -------------------------------------------------------------------------------
\section{Event reconstruction and selection}
\label{sec:event-reco-selection}
% -------------------------------------------------------------------------------

In order to have a dataset that is enriched in events with a hadronically decaying
$W$ boson, and in which the resulting jets can be identified with reasonable accuracy,
this analysis targets the $t\bar{t} \to b\bar{b}W(\to\ell\nu)W(\to
q\bar{q}^{\prime})$ final state, where $\ell$ refers to electrons and
muons.\footnote{Electrons and muons produced via an intermediate $\tau$-lepton
decay are also accepted.} Such a sample provides access to both a pair of
colour-connected ($q\bar{q}^{\prime}$) and non-connected ($b\bar{b}$) jets.

In the following, the definitions used for the object reconstruction, as well as
the event selection used to obtain a signal-enriched sample in data, are
discussed.

\subsection{Detector-level objects}
\label{sec:object-reco-level}

Primary vertices are constructed from all reconstructed tracks
compatible with the interaction region given by LHC beam-spot
characteristics~\cite{ATL-PHYS-PUB-2015-026}.
The hard-scatter primary vertex is then
selected as the vertex with the largest $\sum \pt^2$, where
tracks entering the summation must satisfy $\pt > \SI{0.4}{\GeV}$.

Candidate electrons are reconstructed by matching tracks from the inner detector
to energy deposits in the electromagnetic calorimeter. Electron
identification (ID) relies on a likelihood classifier constructed from various
detector inputs such as calorimeter shower shape or track
quality~\cite{PERF-2013-03,ATLAS-CONF-2014-032,ATL-PHYS-PUB-2015-041}. The electron
candidates must satisfy a ``tight'' ID criterion as defined
in Ref.~\cite{ATL-PHYS-PUB-2015-041}. They must further satisfy $E_{\mathrm{T}} > \SI{25}{\GeV}$ and $\left| \eta
\right| < 2.47$, with the region $1.37 \leq \left| \eta \right| \leq 1.52$ being
excluded. This is the transition region between the
barrel and endcap of the electromagnetic calorimeter, and as a result the energy
resolution is significantly degraded within this region. Isolation
requirements using calorimeter and tracking requirements are applied to reduce
background from non-prompt and fake electrons~\cite{TOPQ-2015-09}. The resulting isolation
efficiency increases linearly with the electron \pt, starting at approximately
90\% and reaching a plateau of 99\% at approximately
$\pt = \SI{60}{\GeV}$. Electrons are also required to have $|d_0^{\text{sig}}| <
5$ and $|z_0 \sin{\theta}| < \SI{0.5}{\mm}$, where $|d_0^{\text{sig}}| =
|d_0|/\sigma_{d_0}$ is the significance of the transverse impact parameter
relative to the beamline, and $z_0$ is the distance along the $z$-axis from the primary vertex
to the point where the track is closest to the beamline.

Muon candidates are reconstructed by matching tracks in the muon
spectrometer to inner-detector tracks. Muons must satisfy a ``medium'' ID
criterion as defined in Ref.~\cite{PERF-2015-10}. The muon \pt is determined
from a fit of all hits associated with the muon track, also taking into
account the energy loss in the calorimeters. Furthermore, muons
must satisfy $\pt > \SI{25}{\GeV}$ and $\left| \eta \right| < 2.5$. Isolation
requirements similar to those used for electrons are applied. The resulting isolation
efficiencies are the same as for electrons. Finally, muon
tracks must have $|d_0^{\text{sig}}| < 3$ and $|z_0 \sin\theta| <
\SI{0.5}{\mm}$.

Jets are reconstructed using the anti-$k_t$ algorithm~\cite{Cacciari:2008gp}
with radius parameter $R=0.4$ as implemented by the
\texttt{FastJet}~\cite{Cacciari:2011ma} package. The inputs to the
jet algorithm consist of three-dimensional, massless, positive-energy
topological clusters~\cite{PERF-2014-07,ATL-PHYS-PUB-2015-036} constructed from
energy deposited in the calorimeters. The jet four-momentum is calibrated
using an $\eta$- and energy-dependent scheme with \textit{in situ} corrections
based on data~\cite{PERF-2012-01,PERF-2016-04}. The calibrated four-momentum is required to
satisfy $\pt > \SI{25}{\GeV}$ and $\left| \eta \right| < 2.5$.
To reduce the number of jets originating from pile-up, an additional selection
criterion based on a jet-vertex tagging technique~\cite{PERF-2014-03} is
applied to jets with $\pt < \SI{60}{\GeV}$ and $\left| \eta \right| < 2.4$.
A multivariate discriminant is used to identify jets containing $b$-hadrons,
using track impact parameters, track invariant mass, track multiplicity and
secondary-vertex information.
The $b$-tagging algorithm~\cite{ATL-PHYS-PUB-2015-022,ATL-PHYS-PUB-2016-012} is
used at a working point that is constructed to operate at an overall
$b$-tagging efficiency of 70\% in simulated \ttbar events for jets
with $\pt > \SI{20}{\GeV}$. The corresponding $c$-jet and light-jet rejection
factors are 12 and 381 respectively, resulting in a purity of 97\%.

Detector information may produce objects that satisfy both the jet and lepton
criteria. In order to match the detector information to a
unique physics object, an overlap removal procedure is
applied: double-counting of electron energy deposits as jets is prevented by
discarding the closest jet lying a distance $\Delta R < 0.2$ from a reconstructed
electron. Subsequently, if an electron lies $\Delta R < 0.4$ from a jet, the
electron is discarded in order to reduce the impact of non-prompt leptons.
Furthermore, if a jet has fewer than three associated tracks and lies
$\Delta R < 0.4$ from a muon, the jet is discarded. Conversely, any muon that lies
$\Delta R < 0.4$ from a jet with at least three associated tracks is discarded.

The magnitude of the missing transverse momentum \met is calculated as the
transverse component of the negative vector sum of the calibrated momentum of
all objects in the event~\cite{PERF-2014-04,ATL-PHYS-PUB-2015-027}. This sum
includes contributions from soft, non-pile-up tracks not associated with any of
the physics objects discussed above.

\subsection{Event selection}
\label{sec:event-selection}

Firstly, basic event-level quality criteria are applied, such as
the presence of a primary vertex and the
requirement of stable detector conditions. Then, events are selected by
requiring that a single-electron or single-muon trigger has fired.
The triggers are designed to select well-identified charged leptons with high
\pt. They require a \pt of at least 20 (26) GeV for muons and 24 (26) GeV for
electrons for the 2015 (2016) data set and also include requirements on the
lepton quality and isolation. These triggers are complemented by triggers with
higher \pt requirements but loosened isolation and identification requirements
to ensure maximum efficiencies at higher lepton \pt.

The reconstructed lepton must satisfy $\pt > \SI{27}{\GeV}$ and must match the
trigger-level object that fired using a geometrical matching. No additional lepton
may be present. Furthermore, selected events must contain at least four jets. At
least two of the jets in the event must be $b$-tagged. Finally,
\met must exceed \SI{20}{\GeV}.

\subsection{Background determination}
\label{sec:background-determination}

After the event selection, a variety of potential background
sources remain. Several sources that contain top quarks contribute to the
background, with events that contain a single top quark being the
dominant contribution. In addition, production of $t\bar{t}+X$ with $X$ being either a $W$, $Z$, or
Higgs boson is an irreducible background, which is, however, expected to be negligible.
Events that contain either two electroweak bosons, or one electroweak boson in
association with jets can be misidentified as signal. However, only the
$W+\text{jets}$ component is expected to contribute significantly.
Finally, multijet processes where either a semileptonic decay of a hadron is wrongly
reconstructed as an isolated lepton or a jet is misidentified as a lepton enter the
signal selection. This last category is collectively called the non-prompt (NP)
and fake lepton background.

All backgrounds are modelled using MC simulation, with the exception of the
NP and fake lepton background, which is estimated using the
matrix method~\cite{TOPQ-2011-02,TOPQ-2012-08}.
A sample enriched in NP and fake leptons is obtained by loosening the requirements on
the standard lepton selections defined in Section~\ref{sec:object-reco-level}.
The efficiency of these ``loose'' leptons to satisfy the standard criteria is then
measured separately for prompt and NP or fake leptons. For both the electrons and muons
the efficiency for a prompt loose lepton to satisfy the standard criteria is
measured using a sample of $Z$ boson decays. The efficiency for NP or fake loose
electrons to satisfy the standard criteria is measured in events with low missing
transverse momentum and the efficiency for NP or fake loose muons to pass the standard
criteria is measured using muons with a high impact parameter significance.
These efficiencies allow the number of NP and fake leptons
selected in the signal region to be estimated.

The number of selected events is listed in Table~\ref{tab:yields}.
The estimated signal purity is approximately 88\%, with the backgrounds from
single top quarks and non-prompt and fake leptons being the largest impurities.
In this analysis, the \ttbar signal includes dilepton \ttbar
  events in which one of the leptons is not identified. These events
  make up 9.8\% of the total \ttbar signal.

\begin{table}[!htb]
  \centering%
  \sisetup{
  table-number-alignment=right,
  round-mode=off}
\begin{tabular}{
  l
  r
  S[table-figures-integer=7, table-figures-decimal=0]
  c
  S[table-figures-integer=5, table-figures-decimal=0]
  }
    \toprule
    Sample & & \multicolumn{3}{c}{Yield} \\
    \midrule
    $t\bar{t}$ & & 1026000 & $\pm$ & 95000 \\
    $t\bar{t}V$ & & 3270 & $\pm$ & 250 \\
    $t\bar{t}H$ & & 1700 & $\pm$ & 100 \\
    Single-top & & 48400 & $\pm$ & 5500 \\
    Diboson & & 1440 & $\pm$ & 220 \\
    $W+\text{jets}$ & & 27700 & $\pm$ & 4700 \\
    $Z+\text{jets}$ & & 8300 & $\pm$ & 1400 \\
    NP/Fake leptons & & 53000 & $\pm$ & 30000 \\
    \midrule
    Total expected & & 1170000 & $\pm$ & 100000 \\
    Observed & & 1153003 &  & \\
\bottomrule
\end{tabular}

  \caption{Event yields after selection. The uncertainties include experimental
    uncertainties and the uncertainty in the data-driven non-prompt and fake
    lepton background. Details of the uncertainties considered can be found in
    Section~\ref{sec:systematics}.}
  \label{tab:yields}
\end{table}

\section{Observable definition and reconstruction}
\label{sec:observables}

The jet-pull vector is calculated from inner-detector tracks created using an
updated reconstruction algorithm~\cite{ATL-PHYS-PUB-2015-051} that makes use of
the newly introduced IBL~\cite{ATLAS-TDR-19} as well as a neural-network-based
clustering algorithm~\cite{PERF-2012-05,PERF-2015-08} 
to improve the pixel cluster position
resolution and the efficiency of reconstructing tracks in jets.
A measurement based on the calorimeter energy clusters of the jet
is not considered in this analysis as it suffers from a significantly degraded
spatial resolution, as was shown in Ref.~\cite{TOPQ-2014-09}.

To ensure good quality, reconstructed tracks must satisfy $\left| \eta \right| <
2.5$ and $\pt > \SI{0.5}{\GeV}$, and further quality cuts are applied to ensure
that they originate from and are assigned to the primary
vertex~\cite{PERF-2015-08}.\footnote{Similar to the quality requirements used
for the electron and muon reconstruction, cuts are applied such that the tracks
satisfy $|d_0| < \SI{2}{\milli\meter}$ and $|z_0\cdot\sin\theta| <
\SI{3}{\milli\meter}$.} This suppresses contributions from pile-up and tracks
with a poor quality fit that are reconstructed from more than one charged
particle.
Matching of tracks to jets is performed using a technique called ghost
association~\cite{Cacciari:2008gn}, in which inner-detector tracks are
included in the jet clustering procedure after having scaled their four-momenta
to have infinitesimal magnitude. As a result, the tracks have no
effect on the jet clustering result whilst being matched to the jet that
most naturally encloses them according to the jet algorithm used. After the
matching procedure, the original track four-momenta are restored.
The jets used in calculating each observable are required to satisfy
$|\eta|<2.1$ so that all associated tracks are within the coverage of the inner
detector. Furthermore, at least two tracks must contribute to the pull-vector
calculation.

The jet axis used to calculate the constituent offsets,
$\vec{\Delta r}_{i}$, in Eq.~\eqref{eq:jet-pull}
is calculated using
the ghost-associated tracks, with their original four-momenta, rather than using the jet axis
calculated from the calorimeter energy clusters that form the jet. This ensures proper
correspondence between the pull vector and the constituents entering its
calculation. For consistency, the total jet \pt in Eq.~\eqref{eq:jet-pull} is also taken
from the four-momentum of the recalculated jet axis.

The analysis presented in this paper measures the colour flow for
two cases:
\begin{enumerate}
\item The signal colour flow is extracted from an explicitly colour-connected dijet
  system.
\item The spurious colour flow is obtained from a jet pair for which no specific
  colour connection is expected.
\end{enumerate}
% the signal colour flow is extracted from an explicitly
% colour-connected dijet system while the spurious colour flow is
% obtained from a jet pair for which no specific colour connection is expected.
Table~\ref{tab:observable-definition} summarises the analysis observables and
their definitions.

\begin{table}[!htb]
  \centering
  \begin{tabular}{c c@{\hskip+0pt} c c}
    \toprule
    \multirow{2}{*}{\textbf{Target colour flow}} && \textbf{Signal colour flow} & \textbf{Spurious colour flow} \\
                                                 && (\small $j_1$ and $j_2$ are colour connected) & (\small $j_1$ and $j_2$ are not colour connected) \\
    \midrule
    % \multirow{2}{*}[-3em]{\textbf{Hard-scatter target}} && {\footnotesize Daughters of hadronically decaying $W$~boson} & {\footnotesize $b$-quarks from top quark decay} \\

    %                                              &&\input{tikz/compact-feynman-ttbar-wHad} & \input{tikz/compact-feynman-ttbar-bbbar} \\
    % \midrule
    \textbf{Jet assignment} && \begingroup
                               \small
                               \renewcommand\arraystretch{1.2}
                               \begin{tabular}{l@{\hskip+0.1em}c@{\hskip+0.1em}l}
                                 $j_1^W$ &:& leading \pt non-$b$-tagged jet \\
                                 $j_2^W$ &:& $2^{\text{nd}}$ leading \pt non-$b$-tagged jet \\
                               \end{tabular} \endgroup & \begingroup
                                                         \small
                                                         \renewcommand\arraystretch{1.2}
                                                         \begin{tabular}{l@{\hskip+0.2em}c@{\hskip+0.2em}l}
                                                           $j_1^b$ &:& leading \pt $b$-tagged jet \\
                                                           $j_2^b$ &:& $2^{\text{nd}}$ leading \pt $b$-tagged jet \\
                                                         \end{tabular} \endgroup \\
    \midrule
    \textbf{Observables} && \begingroup
                            \small
                            \renewcommand\arraystretch{1.2}
                            \begin{tabular}{l@{\hskip+0.2em}c@{\hskip+0.2em}l}
                              $\theta_{\mathcal{P}}\left( j_1^W, j_2^W \right)$ &:& \textit{``forward pull-angle''} \\
                              $\theta_{\mathcal{P}}\left( j_2^W, j_1^W \right)$ &:& \textit{``backward pull-angle''} \\
                              $|\vec{\mathcal{P}}\left( j_1^W \right)|$ &:& \textit{``pull-vector magnitude''} \\
                            \end{tabular}
    \endgroup &  \begingroup
                \small
                \renewcommand\arraystretch{1.2}
                \begin{tabular}{l@{\hskip+0.2em}c@{\hskip+0.2em}l}
                  $\theta_{\mathcal{P}}\left( j_1^b, j_2^b \right)$ &:& \textit{``forward di-$b$-jet-pull angle''} \\
                \end{tabular}
    \endgroup \\
    \bottomrule
  \end{tabular}
  \caption{Summary of the observables' definitions.}
  \label{tab:observable-definition}
\end{table}

The study of the signal colour flow is performed using the candidate
daughters of the
hadronically decaying $W$~boson from the top-quark decay. In practice, the two
leading (highest-\pt) jets that have not been $b$-tagged are selected as
$W$~boson daughter candidates. A dedicated study using simulated $t\bar{t}$
events has shown that this procedure achieves correct matching of both jets in
about 30\% of all events, with roughly 50\% of all cases having a correct match
to one of the two jets. This reduces the sensitivity of this analysis to
different colour model predictions compared with the ideal case of perfect
identification of the $W$~boson daughter jets. Nevertheless, the procedure is
still sufficient to distinguish between the colour models considered in this
analysis.

% The jet assignment procedure
% provides fairly good protection against a non-$W$-jet assignment. Reduced sensitivity
% is observed due to contamination from events where the jet assignment fails to
% provide the correct jets from the $W$ boson.

% The second half of the above paragraph strikes me as important, and so it's important to be clear. 
% The simulation study tells us that in 30 per cent of the events you get the two jets from the 
% hadronically decaying W. In 50 per cent of the events one jet is from the hadronically decaying W
% and the other jet is from the b-system (if b-tagging failed), or..??? 
% In any case I don't understand what is meant by 
% 'good separation from a non-W-jet assignment'? 
% Aren't 50% of events finishing up with a W daghter candidate that does not come from a W?
% (What is the distribution of number of jets in your selected events?)    

\begin{figure}[!htb]
  \centering
  \begin{subfigure}{.49\textwidth}
    \centering
    \includegraphics[width=.99\columnwidth]{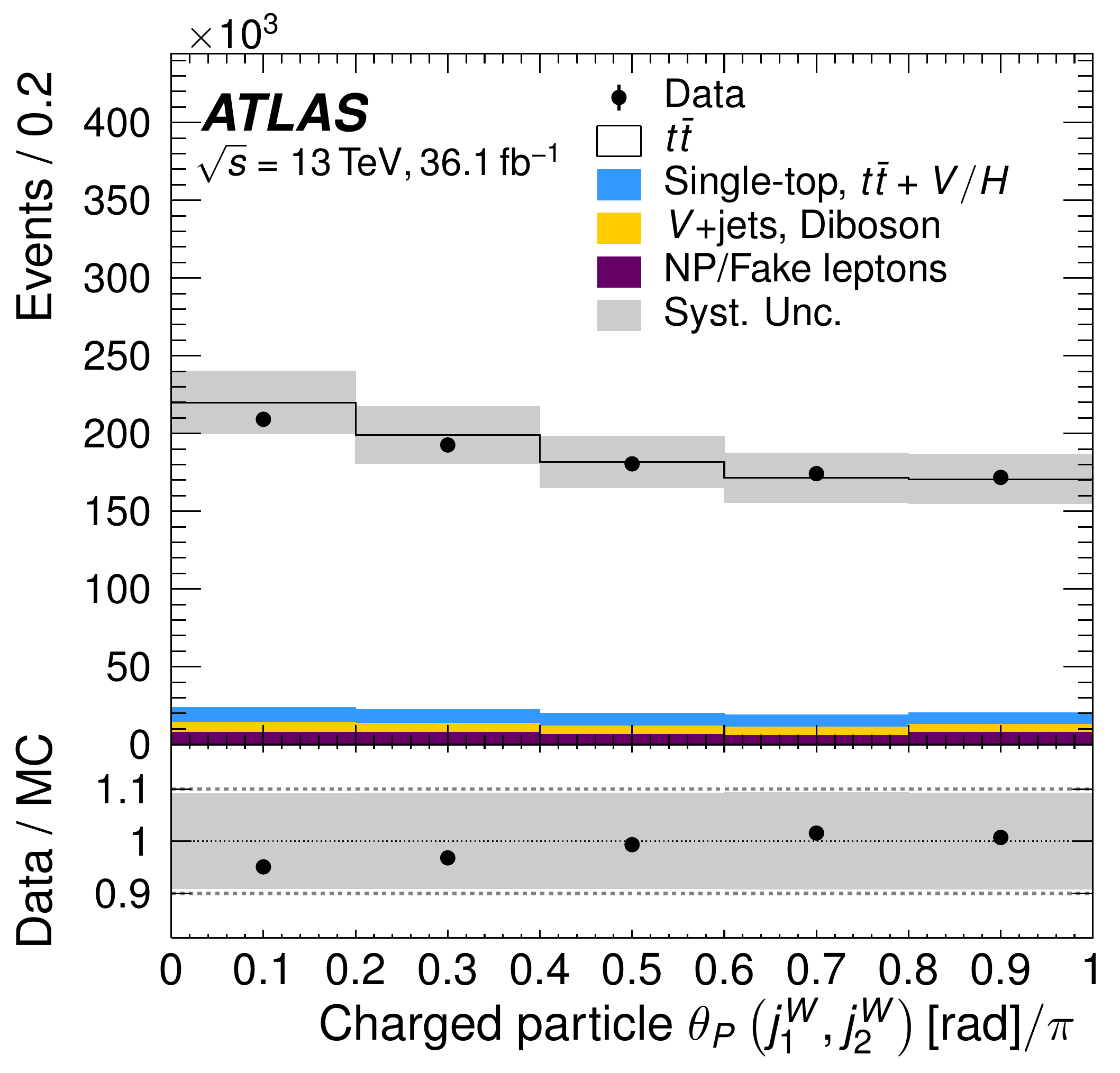}
    \caption{$\theta_{\mathcal{P}}\left( j_1^W, j_2^W \right)$}
    \label{fig:reco-level-Charged-NonBFwd-Angle}
  \end{subfigure}
  \hfill
  \begin{subfigure}{.49\textwidth}
    \centering
    \includegraphics[width=.99\columnwidth]{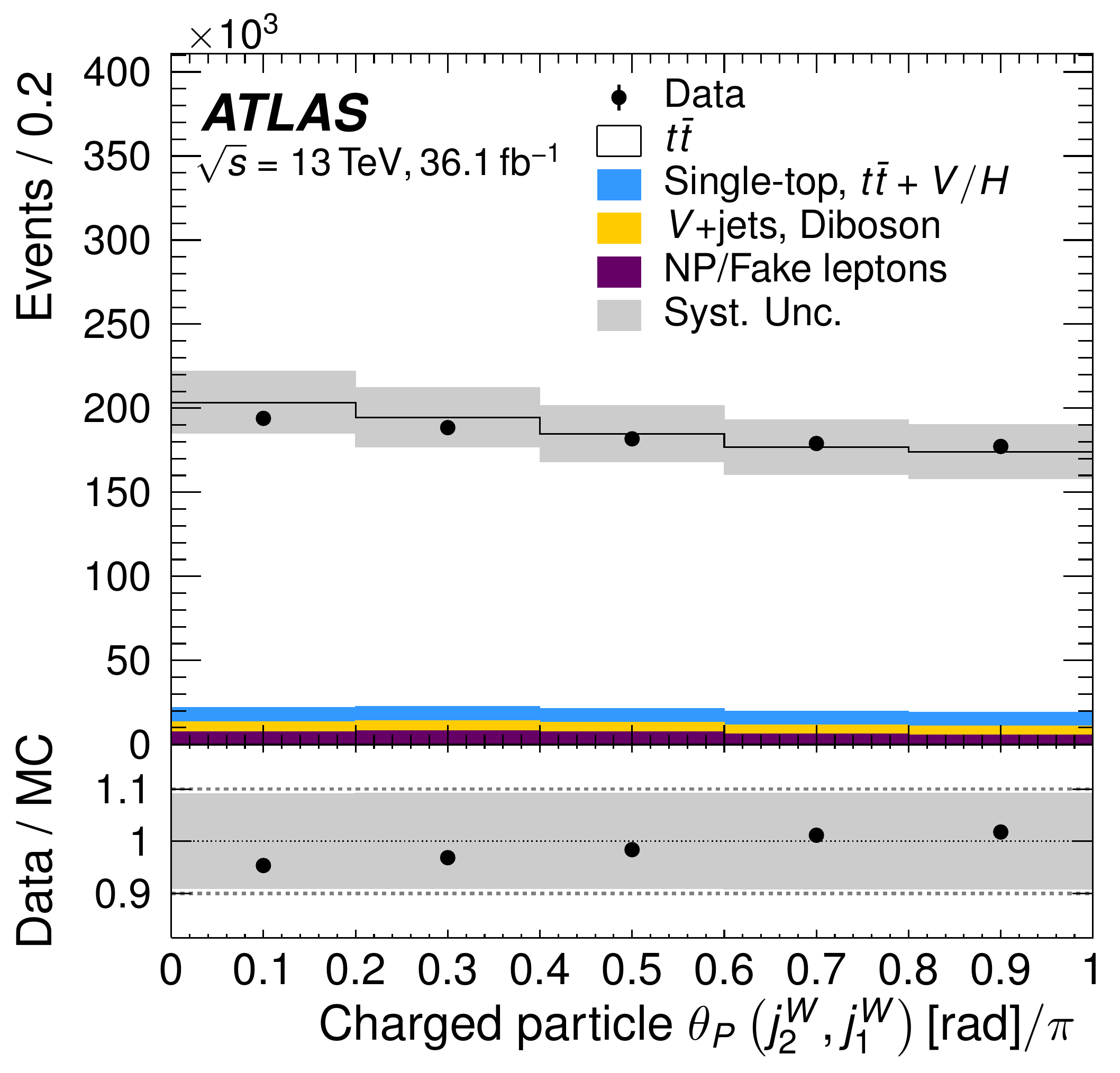}
    \caption{$\theta_{\mathcal{P}}\left( j_2^W, j_1^W \right)$}
    \label{fig:reco-level-Charged-NonBBwd-Angle}
  \end{subfigure}
  \\
  \begin{subfigure}{.49\textwidth}
    \centering
    \includegraphics[width=.99\columnwidth]{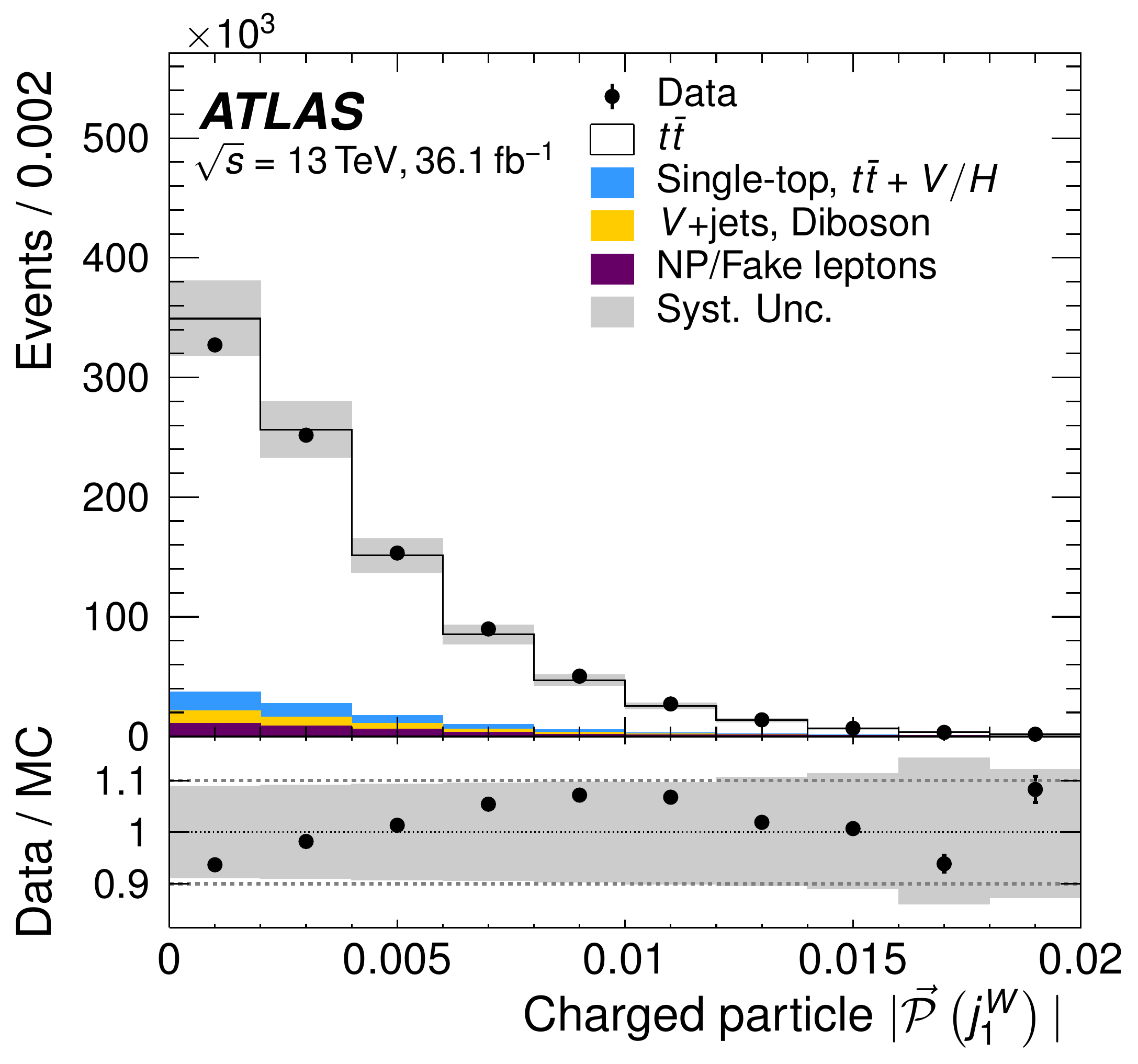}
    \caption{$|\vec{\mathcal{P}}\left( j_1^W \right)|$}
    \label{fig:reco-level-Charged-NonBFwd-Magnitude}
  \end{subfigure}
  \hfill
  \begin{subfigure}{.49\textwidth}
    \centering
    \includegraphics[width=.99\columnwidth]{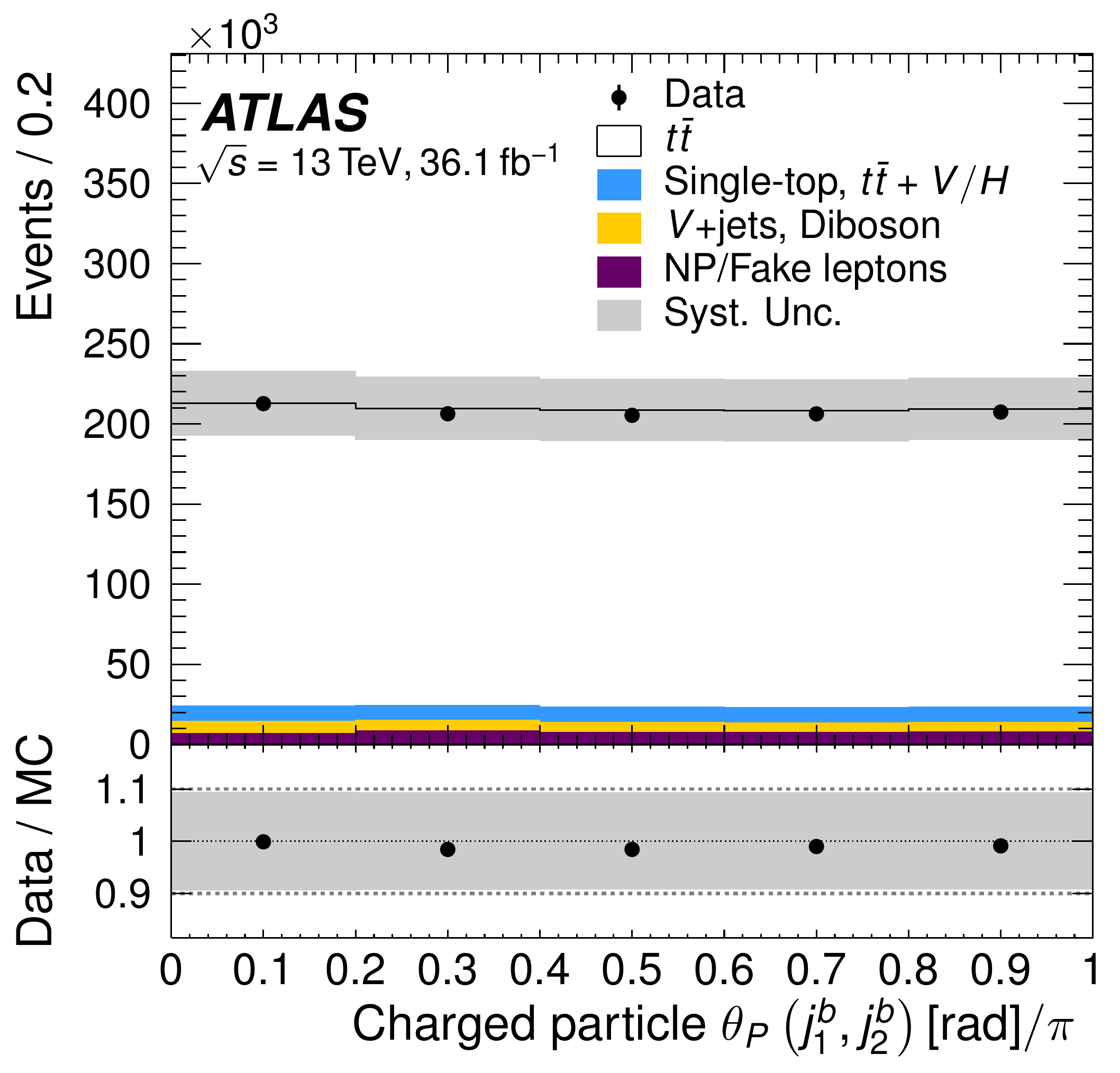}
    \caption{$\theta_{\mathcal{P}}\left( j_1^b, j_2^b \right)$}
    \label{fig:reco-level-Charged-BFwd-Angle}
  \end{subfigure}
  \caption{Detector-level distributions for the four considered observables: the
    \subref{fig:reco-level-Charged-NonBFwd-Angle}~forward and
    \subref{fig:reco-level-Charged-NonBBwd-Angle}~backward pull angle for the
    hadronically decaying $W$~boson daughters,
    \subref{fig:reco-level-Charged-NonBFwd-Magnitude}~the magnitude of the
    leading $W$ daughter's jet-pull vector, and
    \subref{fig:reco-level-Charged-BFwd-Angle} the forward di-$b$-jet-pull
    angle. Uncertainty bands shown include the experimental uncertainties and
    the uncertainty in the data-driven non-prompt and fake lepton background.
    Details of the uncertainties considered can be found in
    Section~\ref{sec:systematics}.}
  \label{fig:reco-level}
\end{figure}

The two jets assigned to the hadronically
decaying $W$~boson are labelled as $j_1^W$ and $j_2^W$, with the
indices referring to their \pt ordering. 
This allows the calculation of two jet-pull angles:
$\theta_{\mathcal{P}}\left( j_1^W, j_2^W \right)$ and
$\theta_{\mathcal{P}}\left(j_2^W, j_1^W \right)$, which are labelled as
``forward pull angle'' and ``backward pull angle'', respectively.
Although the two observables probe the same colour structure, in practice the two values
obtained for a single event have a linear correlation
of less than 1\% in data and can be used for two practically
independent measurements. Figures~\ref{fig:reco-level-Charged-NonBFwd-Angle}
and~\ref{fig:reco-level-Charged-NonBBwd-Angle} compare the distributions
observed for these two pull angles to those predicted by simulation at detector level.

In addition, the magnitude of the jet-pull vector is calculated for the jet with larger
transverse momentum: $|\vec{\mathcal{P}} \left( j_1^W \right)|$. A comparison of the
observed and predicted distributions for this observable can be found in
Figure~\ref{fig:reco-level-Charged-NonBFwd-Magnitude}, which shows a steeply
falling distribution largely contained in the region
below $0.005$.

In \ttbar events an obvious candidate for measuring spurious colour flow is the structure
observed between the two leading $b$-tagged jets,
as the partons that initiate the $b$-jets
are not expected to have any specific colour connection. For a typical signal
event, their colour charge can be traced to the gluon that splits into the
$t\bar{t}$ pair. This coloured initial state ensures that the two $b$-quarks are not
expected to be colour connected. Therefore, the forward di-$b$-jet-pull angle is
calculated from the two leading $b$-tagged jets: $\theta_{\mathcal{P}}\left(
j_1^b, j_2^b \right)$. According to the \ttbar simulation, this choice achieves correct 
matching for both jets in about 80\% of all events.
Figure~\ref{fig:reco-level-Charged-BFwd-Angle} shows a comparison of the
distribution observed in data to that predicted by simulation for this
observable. Consistent with the expectation, the distribution is flat, unlike in
the case of the jet pairs from $W$~boson decays.

% \input{sections/particle-level}
% -------------------------------------------------------------------------------
\section{Unfolding}
\label{sec:unfolding}
% -------------------------------------------------------------------------------

Particle-level objects are selected in simulated events using definitions analogous
to those used at detector level, as discussed in the previous section.
Particle-level objects are defined using
particles with mean lifetime greater than $\SI{30}{\pico\second}$.

Electrons and muons must not originate from a hadron in the MC
generator-level event record, either directly or through an intermediate $\tau$-lepton decay.
In effect, this means that the lepton originates from a
real $W$ or $Z$ boson. To take into account final-state photon radiation,
the lepton four-momentum is modified by adding to it all photons
not originating from a hadron that are within a $\Delta R = 0.1$ cone around
the lepton. Leptons are then required to satisfy $\pt > \SI{25}{\GeV}$ and
$\left| \eta \right| < 2.5$.

Particle-level jets are constructed by clustering all stable particles, excluding leptons not
from hadron decays and their radiated photons,
using the same clustering algorithm and configuration as is used for the
detector-level jets. Particle-level jets are furthermore required to
satisfy $\pt > \SI{25}{\GeV}$ and $\left| \eta \right| < 2.5$.
Classification of jets as having originated from a $b$-hadron is
performed using ghost association~\cite{Cacciari:2008gn} where the $b$-hadrons
considered for the procedure must satisfy $\pt > \SI{5}{\GeV}$. This is
equivalent to the method used for matching tracks to jets described in
Section~\ref{sec:observables},
except that it is applied during particle-level jet clustering and adds ghosts
for unstable $b$-hadrons rather than inner-detector tracks. A particle-level
jet is considered to be $b$-tagged if it contains at least one such $b$-hadron.

An overlap removal procedure is applied that rejects leptons that overlap
geometrically with a jet at $\Delta R < 0.4$.

The magnitude of the missing transverse momentum \met at
particle level is calculated as the transverse component of the four-momentum
sum of all neutrinos in the event excluding those from hadron decays,
either directly or through an intermediate $\tau$-lepton decay.

At particle level, the event selection requires exactly one
lepton with $\pt > \SI{27}{\GeV}$ with no additional lepton, at least four jets
of which at least two are $b$-tagged, as well as $\met > \SI{20}{\GeV}$.

At particle level, the input to the calculation of the jet-pull vector is
the collection of jet constituents as defined by the clustering procedure
described in Section~\ref{sec:object-reco-level}. To reflect the fact that the
detector-level observable's definition uses tracks, only charged particles are considered.
Furthermore, a requirement of $\pt >
\SI{0.5}{\GeV}$ is imposed in line with the detector-level definition
to reduce simulation-based extrapolation and associated uncertainties.
Apart from the inputs to the jet-pull-vector calculation, the procedure applied
at detector level is mirrored exactly at particle level.

The measured distributions are unfolded using
the iterative Bayesian method~\cite{2010arXiv1010.0632D} as implemented by
the \texttt{RooUnfold} framework~\cite{Adye:2011gm}. This algorithm
iteratively corrects the observed data to an unfolded
particle-level distribution given a certain particle-level prior. Initially, this
prior is taken to be the particle-level distribution obtained from simulation.
However, it is updated after each iteration with the observed posterior
distribution. Thus, the algorithm converges to an unfolded result driven by
the observed distribution.

The measurement procedure consists essentially of two stages: first the
background contributions are subtracted bin-by-bin from the observed data.
Secondly, detector effects are unfolded from the signal
distribution using a detector response model, the migration matrix, obtained
from simulated $t\bar{t}$ events. As
part of this second step, two correction factors are applied that correct for
non-overlap of the fiducial phase space at detector- and particle-level.
The corrections account for events that fall within the fiducial phase space of
one level but not the other. The full procedure for an observable
$X$ can be summarised symbolically by the equation

\begin{align*}
  % D. Stoker marked this as non-numbered. Previously we've been asked to include numbers even on non-referenced equations.
  \frac{\mathrm{d} \sigma_{\text{Fid}}^{t}}{\mathrm{d} X^{t}} = \frac1{\mathcal{L} \cdot \Delta X^{t}} \cdot \frac1{\epsilon^t} \sum_{r} \mathcal{M}_{r,t}^{-1} \cdot \epsilon_{\text{Fid}}^{r} \cdot \left( N_{\text{Obs}}^{r} - N_{\text{Bkg}}^{r} \right) \, ,
  % \label{eq:unfolding}
\end{align*}

where $t$ indicates the particle-level bin index, $r$ the detector-level bin
index, $\mathcal{L}$ is the
integrated luminosity of the data, $\mathcal{M}_{r,t}$ is the migration matrix
and the inversion symbolises unfolding using the iterative Bayesian method,
$N_{\text{Obs}}^{r}$ is the number of observed events, $N_{\text{Bkg}}^{r}$ the
expected number of background events, and $\epsilon^t$ and $\epsilon_{\text{Fid}}^r$ are
the phase-space correction factors. These last two parameters are defined as

\begin{align*}
  % D. Stoker marked this as non-numbered. Previously we've been asked to include numbers even on non-referenced equations.
  \epsilon^t = \frac{N_{\text{PL}\wedge\text{RL}}}{N_{\text{PL}}} \qquad\qquad\qquad\qquad \epsilon^r_{\text{Fid}} = \frac{N_{\text{PL}\wedge\text{RL}}}{N_{\text{RL}}} \, .
\end{align*}

The number $N_{\text{PL}}$ ($N_{\text{RL}}$) indicates the number of events fulfilling the
fiducial requirements at particle level (selection requirements at
detector level), $N_{\text{PL}\wedge\text{RL}}$ is the number of events that
pass both sets of requirements at their respective level.

The response model and phase-space correction factors are obtained from
$t\bar{t}$ simulation.

Some of the background samples considered in this analysis potentially contain
true signal colour flow, e.g.\ the single-top or $t\bar{t}+X$ contributions.
However, as their overall contributions are very small, even extreme changes in
their respective colour flow have a negligible effect. Therefore, all such
contributions are ignored and the estimated backgrounds, with SM colour flow
assumed, are subtracted from the data.

The binning chosen for the observables is determined by optimisation studies
performed with simulated samples. A good binning choice should result in a
mostly diagonal migration matrix with bin widths appropriate to the observed
resolution. The optimisation therefore imposes a requirement of having at least
50\% of events on-diagonal for each particle-level bin of the migration
matrix.

The number of iterations used by the unfolding method is
chosen such that the total uncertainty composed of the statistical
uncertainty and the bias is minimised.

% -------------------------------------------------------------------------------
\section{Treatment of uncertainties}
\label{sec:systematics}
% -------------------------------------------------------------------------------

Several systematic uncertainties affect the measurements discussed
above. The different sources are grouped into four categories:
experimental uncertainties, uncertainties related to the modelling of the
signal process, uncertainties related to the modelling of the
background predictions, and an uncertainty related to the unfolding procedure.

The changes that result from variations accounting for sources of systematic uncertainty are used to calculate a
covariance matrix for each source individually. This covariance matrix combines
the changes from all measured observables simultaneously, and therefore
also includes the cross-correlations between observables. 
The total covariance
matrix is then calculated by summation over the covariances obtained from
all sources of systematic uncertainty. The changes observed for a source of systematic uncertainty are
symmetrised prior to calculating the covariance. For one-sided variations, the
change is taken as a symmetric uncertainty. For two-sided variations, which
variation is used to infer the sign is completely arbitrary, as long as it is
done consistently. In this analysis, the sign~---~which is only relevant for the
off-diagonal elements of the covariance matrix~---~is taken from the upward
variation while the value is taken as the larger change. Furthermore, it is
assumed that all uncertainties, including modelling uncertainties, are
Gaussian-distributed.

\subsection{Experimental uncertainties}
\label{sec:systematics-experiment}

Systematic uncertainties due to the modelling of the detector response and other
experimental sources affect the signal reconstruction efficiency, the unfolding
procedure, and the background estimate. Each source of experimental uncertainty
is treated individually by repeating the full unfolding procedure using
as input a detector response that has been varied appropriately. 
The unfolding result is then compared to the nominal result
and the difference is taken as the systematic uncertainty.
Through this procedure the measured data enter the calculation for each source of
experimental uncertainty.  

Uncertainties due to lepton identification, isolation, reconstruction, and
trigger requirements are evaluated by varying the scale factors applied in the simulation
to efficiencies and kinematic calibrations within their uncertainties.
% from the values that gave agreement between data and simulation.
The scale factors and an estimate of their uncertainty 
were derived from data in
control regions enriched in $Z \to \ell\ell, W \to \ell\nu$, or $J/\psi$ events
~\cite{PERF-2010-04,PERF-2015-10,ATL-PHYS-PUB-2016-015,ATLAS-CONF-2016-024}.

The uncertainties due to the jet energy scale (JES) and resolution (JER) are
derived using a combination of simulation, test-beam data, and \textit{in situ}
measurements~\cite{PERF-2011-03,PERF-2011-04,PERF-2011-05,PERF-2012-01,ATL-PHYS-PUB-2015-015}.
In addition, contributions from $\eta$-intercalibration, single-particle
response, pile-up, jet flavour composition, punch-through, and varying calorimeter
response to different jet flavours are taken into account. This results in a
scheme with variations for 20 systematic uncertainty contributions to the JES.

Efficiencies related to the performance of the $b$-tagging procedure are
corrected in simulation to account for differences between data and simulation.
The corresponding scale factors are extracted from simulated \ttbar events.
This is done separately for $b$-jets, $c$-jets, and light jets, thereby accounting for
mis-tags. Uncertainties related to this procedure are propagated by varying the scale
factors within their
uncertainty~\cite{ATLAS-CONF-2014-004,ATLAS-CONF-2014-046,ATL-PHYS-PUB-2015-022}.

The uncertainties on the \met due to systematic shifts in the
corrections for leptons and jets are accounted for in a fully correlated way in
their evaluation for those physics objects. Uncertainties due to track-based
terms in the \met calculation, i.e.\ those that are not
associated with any other reconstructed object, are treated
separately~\cite{PERF-2011-07}.

All uncertainties associated with the reconstructed tracks directly enter the observable
calculation as defined in Eq.~\eqref{eq:jet-pull}. Uncertainties are either
expressed as a change in the tracking efficiency or smearing of the track
momentum~\cite{ATL-PHYS-PUB-2015-051,PERF-2015-08}. This also includes effects
due to fake tracks and lost tracks in the core of jets. Corrections and scale
factors were extracted using simulated data as well as experimental data
obtained from minimum-bias, dijet, and $Z \to \mu\mu$ selections. The systematic shifts
applied as part of this procedure are in most cases parameterised as functions
of the track \pt and $\eta$, see Ref.~\cite{ATL-PHYS-PUB-2015-051}.

The uncertainty in the combined 2015 and 2016 integrated luminosity is
2.1\%, which is derived following a method similar to that
detailed in Ref.~\cite{DAPR-2013-01}, from a calibration of the luminosity scale
using $x$--$y$ beam-separation scans performed in August 2015 and May 2016. This uncertainty
affects the scaling of the background prediction that is subtracted from the
observed data. The uncertainty related to the pile-up reweighting is evaluated
by varying the scale factors by their uncertainty based on the reweighting of
the average number of interactions per $pp$ collision.

The data's statistical uncertainty and bin-to-bin correlations are evaluated using
the bootstrap method~\cite{efron1979}. Bootstrap
replicas of the measured data are propagated through the unfolding procedure and
their variance is used to assess the statistical uncertainty. These replicas
can also be used to calculate the statistical component of the covariance of the
measurement as well as the statistical bin-by-bin correlations of the pre- or
post-unfolding distributions.

\subsection{Signal modelling uncertainties}
\label{sec:systematics-signal-model}

The following systematic uncertainties related to the modelling of the
$t\bar{t}$ system are considered: the choice of
matrix-element generator, the choice of PDF, the hadronisation model, the
amount of initial- and final-state radiation (ISR/FSR), and the amount and
strength of colour reconnection (CR).

Signal modelling uncertainties are evaluated individually using different
signal MC samples. Detector-level
distributions from the alternative signal MC sample are unfolded using the nominal
response model. The unfolding result is then compared to the particle-level
prediction of the alternative MC sample and the difference is used as the uncertainty.
Table~\ref{tab:mc-samples} lists the alternative signal MC samples used for
assessing the generator, hadronisation, ISR/FSR systematic uncertainties (\texttt{A14.v3c}
tune variations), and CR (\texttt{A14.v1} tune variations) systematic uncertainties.

The uncertainty arising from the choice of PDF is evaluated by
creating reweighted pseudo-samples, in which the weight variations for
the PDF sets are according to the
\texttt{PDF4LHC}~\cite{Butterworth:2015oua} prescription. 
The unfolding
results obtained for the pseudo-samples are then combined in accordance with the
\texttt{PDF4LHC} procedure to obtain a single systematic shift.

\subsection{Background modelling uncertainties}
\label{sec:systematics-background-model}

Systematic uncertainties related to the background modelling affect the number
of background events subtracted from data prior to the unfolding.

The normalisation of the background contributions obtained from MC simulation is
varied within the uncertainties obtained from the corresponding cross-section
calculation. For the single-top background, the normalisation uncertainty ranges
from 3.6\% to 5.3\%~\cite{PhysRevD.81.054028,PhysRevD.82.054018,PhysRevD.83.091503}, and for the $t\bar{t}Z$ and $t\bar{t}W$
backgrounds it is 12\% and 13\%, respectively~\cite{Alwall:2014hca, deFlorian:2016spz}. In the case of the
$W/Z+\text{jets}$ backgrounds, the uncertainties include a contribution from the
overall cross-section normalisation (4\%), as well as an additional
24\% uncertainty added in quadrature for each
jet~\cite{ATLAS-CONF-2015-039,BERENDS199132}. For the diboson background, the
normalisation uncertainty is 6\%~\cite{ATL-PHYS-PUB-2016-002}. The
uncertainty of the normalisation for the $t\bar{t}H$ background is chosen to be
100\%.

The uncertainty arising from the modelling of the non-prompt and fake lepton
background is assessed by varying the normalisation by 50\%, as well
as by changing the efficiency parameterisation used by the matrix method~\cite{TOPQ-2011-02,TOPQ-2012-08} to
obtain a shape uncertainty. These uncertainties were found to cover adequately
any disagreement between data and prediction in various background-dominated
control regions.

The uncertainty due to the level of radiation in the single-top
background is evaluated using two alternative simulation samples with varied
levels of radiation. These two samples were generated
using the same approach that was used to produce the radiation variation
samples of the nominal \ttbar process. The uncertainty due to the higher-order
overlap between the \ttbar and $Wt$ processes is evaluated by assessing the
impact of replacing the nominal $Wt$ MC sample, which accounts for overlap
using the ``diagram removal'' scheme, with an alternative $Wt$ MC sample that
accounts for the overlap using the ``diagram subtraction''
scheme~\cite{2008JHEP...07..029F}.

A $Wt$ colour-model uncertainty is considered, which is motivated by the
overlap between the \ttbar and $Wt$ processes. This overlap implies that the
colour flow in $Wt$ is of the same type as the signal colour flow in the \ttbar
process. However, the $Wt$ colour flow is estimated from simulation and
subtracted from data prior to unfolding. Hence, mismodelling of the $Wt$ colour
flow would affect the unfolding result. An uncertainty is constructed by
reweighting the combination of \ttbar and $Wt$ to have the same shape
as data.
For evaluation of the systematic uncertainty, the reweighted $Wt$ is then considered for the
background subtraction and unfolding is repeated.

\subsection{Unfolding procedure systematic uncertainty}
\label{sec:systematics-unfolding-procedure}

The uncertainty arising from the unfolding procedure, also called the non-closure
uncertainty, is assessed using a data-driven approach. For each measured
distribution, simulated particle-level events are reweighted using a linear weight
function such that the corresponding detector-level distributions are in better
agreement with the data. The weights are propagated to the corresponding
detector-level events and the resulting distributions are unfolded using the
nominal detector-response model. Deviations of these unfolded distributions from
the reweighted particle-level distributions are then assigned as the non-closure uncertainty.

A summary of the uncertainties affecting $\theta_{\mathcal{P}}\left( j_1^W,
  j_2^W \right)$ is shown in Table~\ref{tab:theta_j1j2_syst_summary}. The total
uncertainty is dominated by systematic uncertainties, with those accounting for
\ttbar modelling being dominant in most bins. Uncertainties that directly affect
the inputs to the pull-vector calculation, such as the JES, JER and track
uncertainties are generally sub-dominant.

The systematic uncertainties in Table \ref{tab:theta_j1j2_syst_summary} are much smaller 
than those shown in Table~\ref{tab:yields} and Figure~\ref{fig:reco-level}. 
This is because Table~\ref{tab:theta_j1j2_syst_summary} gives the uncertainties appropriate for 
a comparison between normalised distributions in which overall scale
uncertainties play no role. As a result, many of the experimental uncertainties,
which have little to no impact on the shape of the measured distributions, also
have a reduced effect on the measurement. For example, the uncertainties due to
$b$-tagging reduce from around 7.5\% to less than 0.5\%.

% This is because Table~\ref{tab:theta_j1j2_syst_summary} gives the uncertainties appropriate for 
% a comparison between normalised distributions, as discussed in Section \ref{sec:result},
% in which overall scale uncertainties play no role.     

\begin{table}[!htb]
  \centering
  \sisetup{
  table-number-alignment=center,
  round-mode=off}
\begin{tabular}{l*{4}S[table-figures-integer=1, table-figures-decimal=2]}
  \toprule
  \multirow{2}{*}{$\Delta \theta_P\left( j_{1}^{W}, j_{2}^{W} \right) \, [\si{\percent}]$} & \multicolumn{4}{c}{$\theta_P\left( j_{1}^{W}, j_{2}^{W} \right)$} \\
  \cmidrule(lr{.75em}){2-5}
  & \multicolumn{1}{c}{$0.0$ -- $0.21$} & \multicolumn{1}{c}{$0.21$ -- $0.48$} & \multicolumn{1}{c}{$0.48$ -- $0.78$} & \multicolumn{1}{c}{$0.78$ -- $1.0$} \\
  \midrule
  Hadronisation & 0.55 & 0.13 & 0.24 & 0.14 \\
  Generator & 0.32 & 0.25 & 0.50 & 0.01 \\
  $b$-tagging & 0.35 & 0.13 & 0.20 & 0.31 \\
  Background model & 0.30 & 0.16 & 0.16 & 0.27 \\
  Colour reconnection & 0.22 & 0.16 & 0.16 & 0.18 \\
  JER & 0.11 & 0.12 & 0.23 & 0.02 \\
  Pile-up & 0.19 & 0.16 & 0.00 & 0.01 \\
  Non-closure & 0.14 & 0.07 & 0.07 & 0.18 \\
  JES & 0.12 & 0.06 & 0.14 & 0.06 \\
  ISR / FSR & 0.15 & 0.02 & 0.12 & 0.02 \\
  Tracks & 0.05 & 0.04 & 0.03 & 0.06 \\
  Other & 0.02 & 0.01 & 0.01 & 0.02 \\
  \midrule
  Syst. & 0.88 & 0.44 & 0.71 & 0.51 \\
  Stat. & 0.23 & 0.19 & 0.19 & 0.25 \\
  Total & 0.91 & 0.48 & 0.73 & 0.57 \\
  \bottomrule
\end{tabular}

  \caption{Statistical and systematic uncertainties affecting the measurement of
    $\theta_{\mathcal{P}} \left( j_1^W, j_2^W \right)$. The category ``Other''
summarises various smaller uncertainty components. Uncertainties are ordered by
the mean value of the uncertainty across all bins and are
    expressed in percent of the measured value.}
  \label{tab:theta_j1j2_syst_summary}
\end{table}

% -------------------------------------------------------------------------------
\section{Results}
\label{sec:result}
% -------------------------------------------------------------------------------

\begin{figure}[!tbh]
  \centering
  \begin{subfigure}{.49\textwidth}
    \centering
    \includegraphics[width=.99\textwidth]{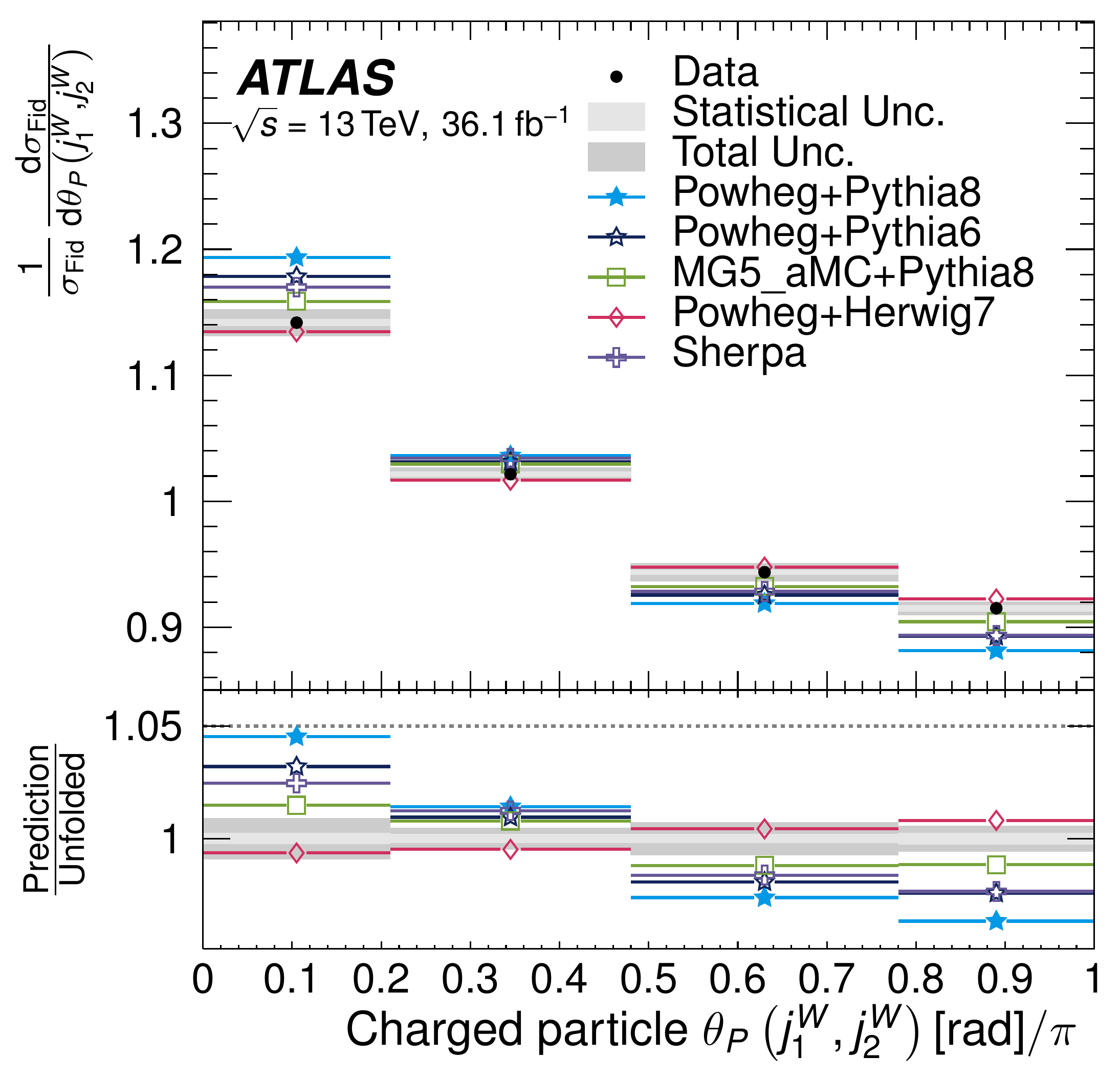}
    \caption{$\theta_{\mathcal{P}}\left( j_1^W, j_2^W \right)$}
    \label{fig:results-NonBFwd-Angle}
  \end{subfigure}
  \hfill
  \begin{subfigure}{.49\textwidth}
    \centering
    \includegraphics[width=.99\textwidth]{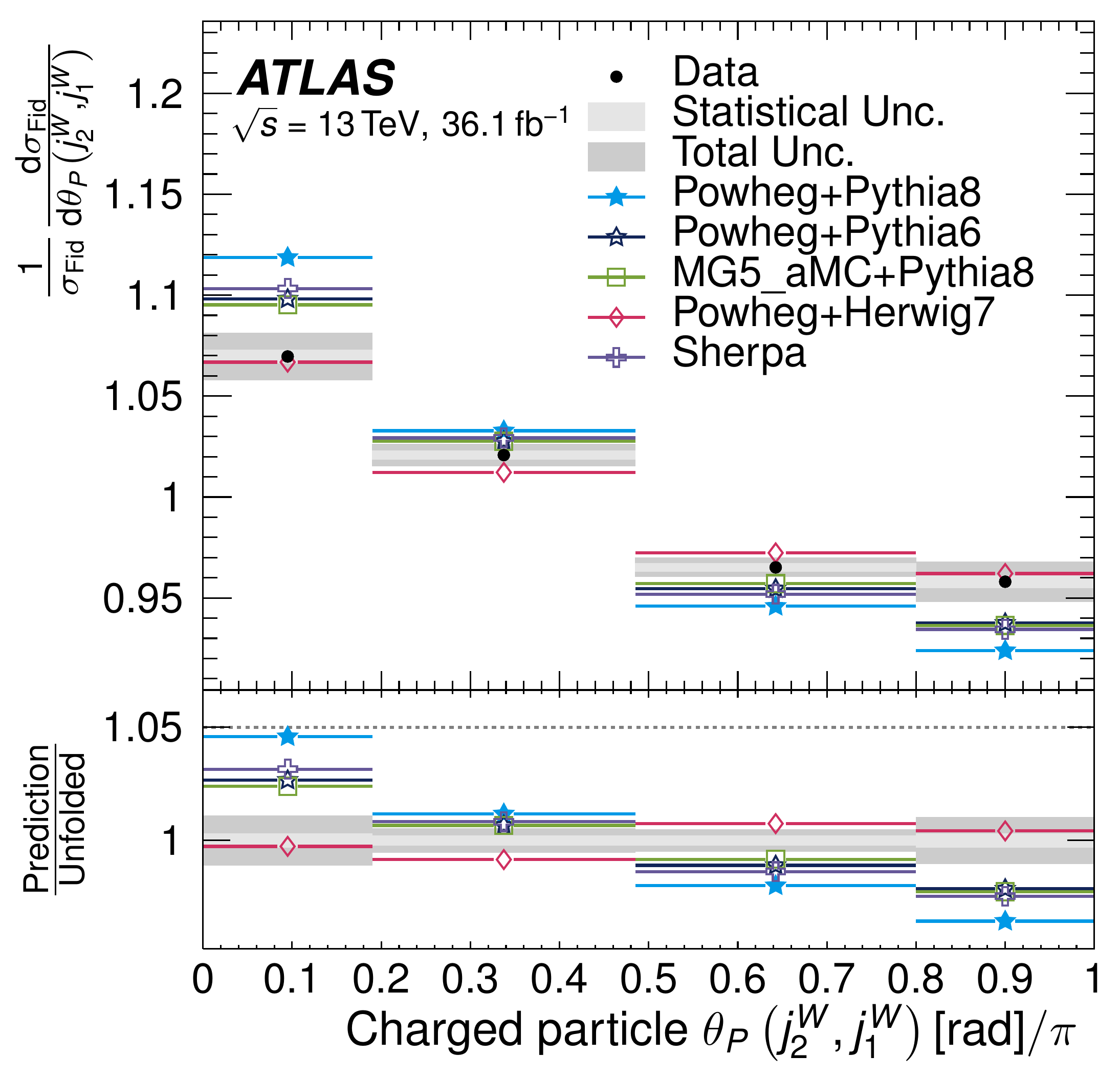}
    \caption{$\theta_{\mathcal{P}}\left( j_2^W, j_1^W \right)$}
    \label{fig:results-NonBBwd-Angle}
  \end{subfigure}
  \\
  \begin{subfigure}{.49\textwidth}
    \centering
    \includegraphics[width=.99\textwidth]{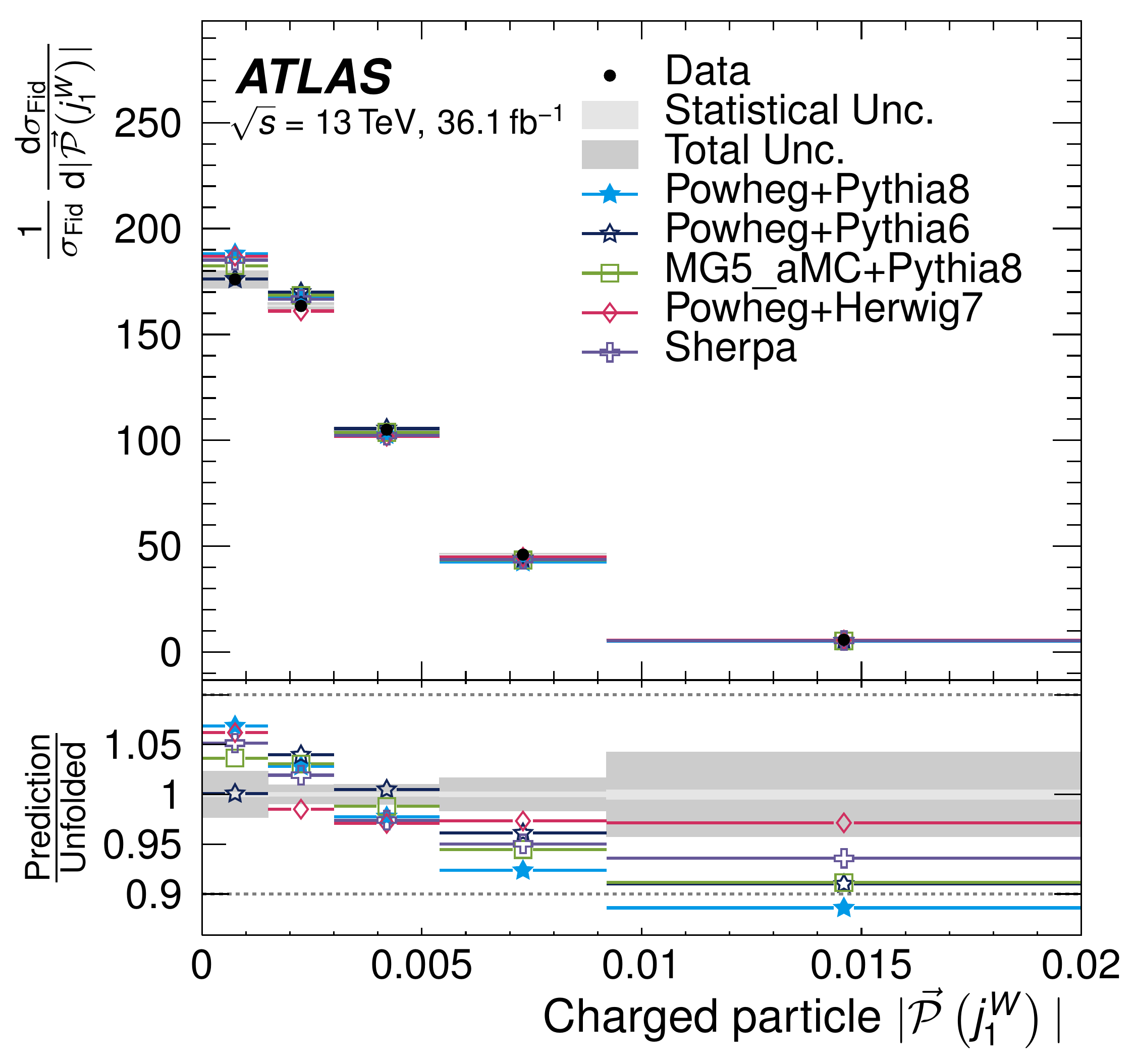}
    \caption{$|\vec{\mathcal{P}}\left( j_1^W \right)|$}
    \label{fig:results-NonBFwd-Magnitude}
  \end{subfigure}
  \hfill
  \begin{subfigure}{.49\textwidth}
    \centering
    \includegraphics[width=.99\textwidth]{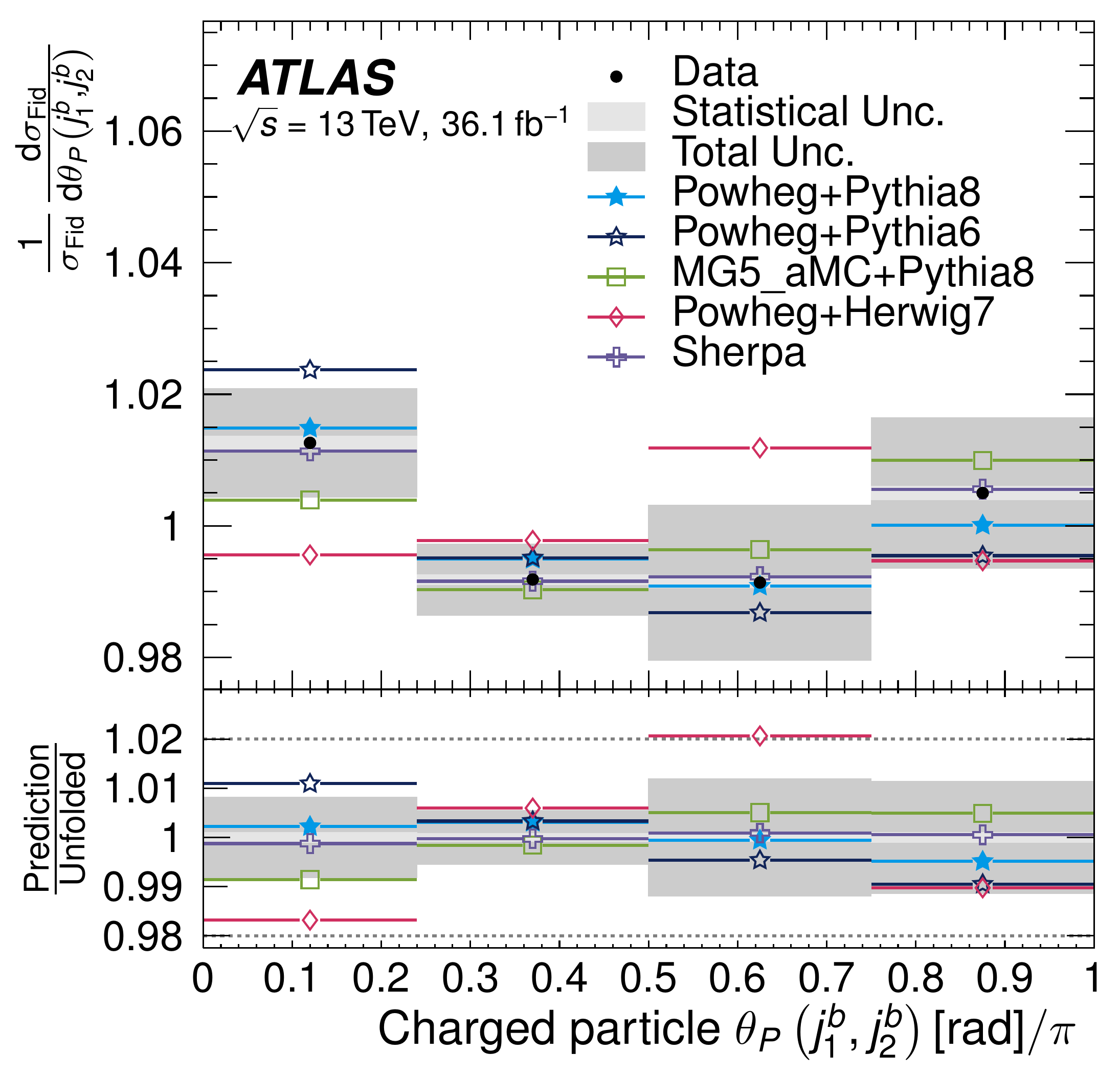}
    \caption{$\theta_{\mathcal{P}}\left( j_1^b, j_2^b \right)$}
    \label{fig:results-BFwd-Angle}
  \end{subfigure}
  \caption{Normalised fiducial differential cross-sections as a function of the
    \subref{fig:results-NonBFwd-Angle}~forward and
    \subref{fig:results-NonBBwd-Angle}~backward pull angle for the
    hadronically decaying $W$ boson daughters,
    \subref{fig:results-NonBFwd-Magnitude} the magnitude of the
    leading $W$ daughter's jet-pull vector, and
    \subref{fig:results-BFwd-Angle} the forward di-$b$-jet-pull
    angle.
    The data are compared to various SM predictions. The statistical
    uncertainties in the predictions are smaller than the marker size.}
  \label{fig:results}
\end{figure}

Figure~\ref{fig:results} compares the normalised unfolded data to several Standard Model (SM) predictions
for all four observables. Three SM predictions use \POWHEG to generate the
hard-scatter events and then differ for the subsequent hadronisation, namely
\PYTHIAV{6}, \PYTHIAV{8}, and \HERWIGV{7}. A main difference between these
predictions is that the \PYTHIA family uses the colour string
model~\cite{Andersson:1983ia} while \HERWIG uses the cluster
model~\cite{Bellm:2015jjp} for hadronisation. One SM prediction uses MG5\_aMC
to produce the hard-scatter event, the hadronisation is then performed using
\PYTHIAV{8}. Finally, one SM prediction is obtained from events generated with
\SHERPA.

\begin{figure}[!tbh]
  \centering
  \begin{subfigure}{.49\textwidth}
    \centering
    \includegraphics[width=.99\textwidth]{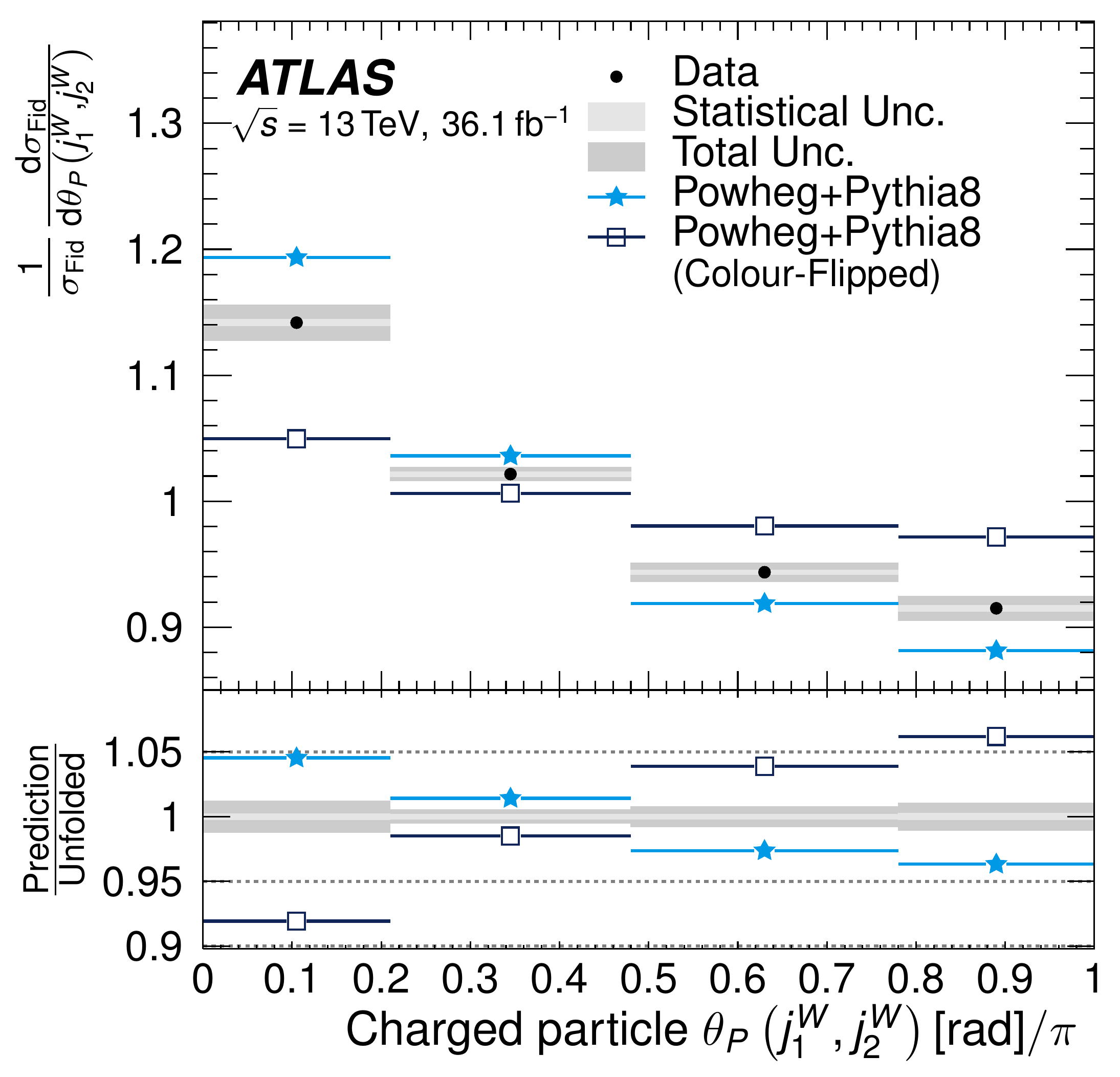}
    \caption{$\theta_{\mathcal{P}}\left( j_1^W, j_2^W \right)$}
    \label{fig:results-CFlip-NonBFwd-Angle}
  \end{subfigure}
  \hfill
  \begin{subfigure}{.49\textwidth}
    \centering
    \includegraphics[width=.99\textwidth]{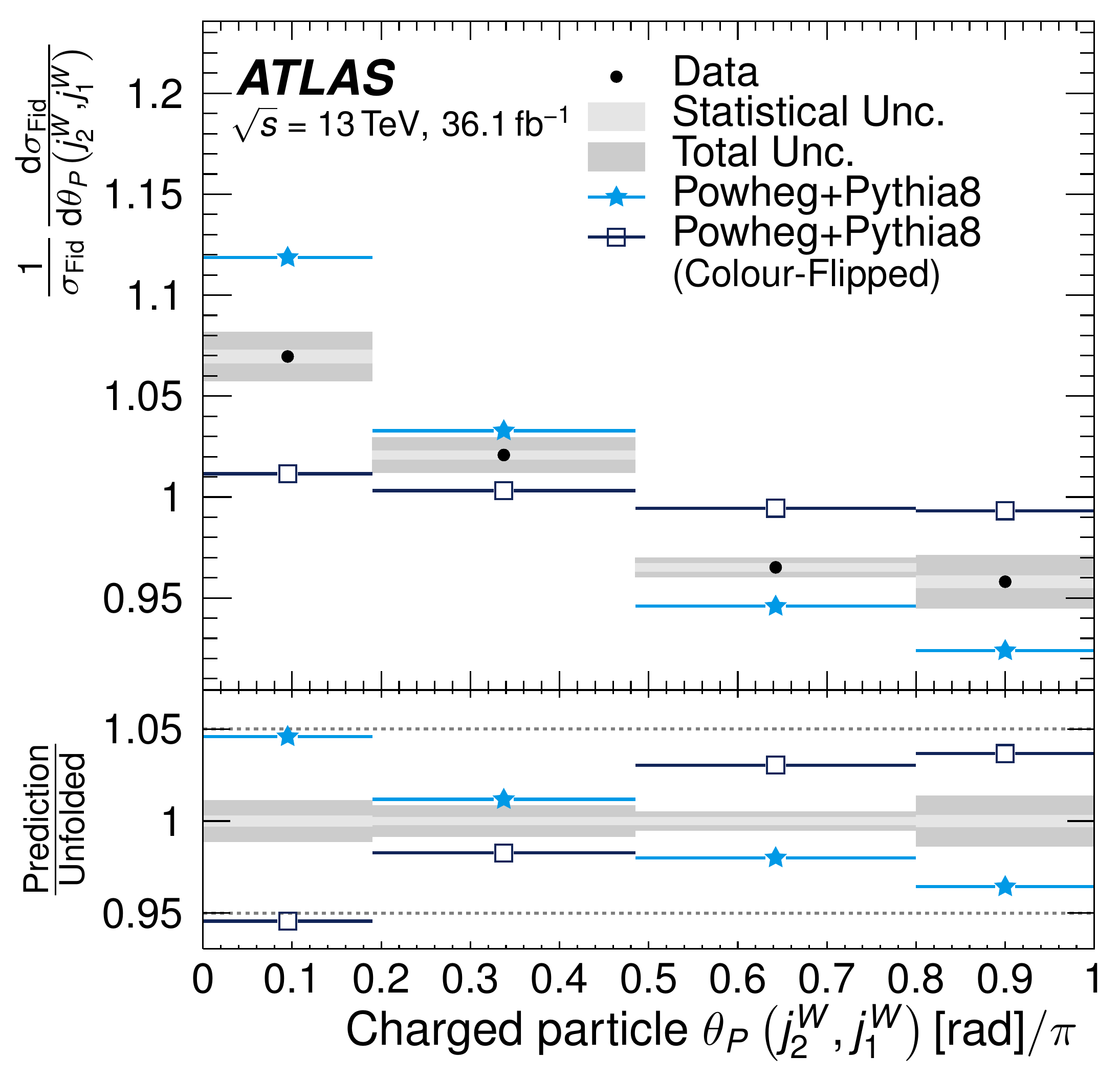}
    \caption{$\theta_{\mathcal{P}}\left( j_2^W, j_1^W \right)$}
    \label{fig:results-CFlip-NonBBwd-Angle}
  \end{subfigure}
  \\
    \begin{subfigure}{.49\textwidth}
    \centering
    \includegraphics[width=.99\textwidth]{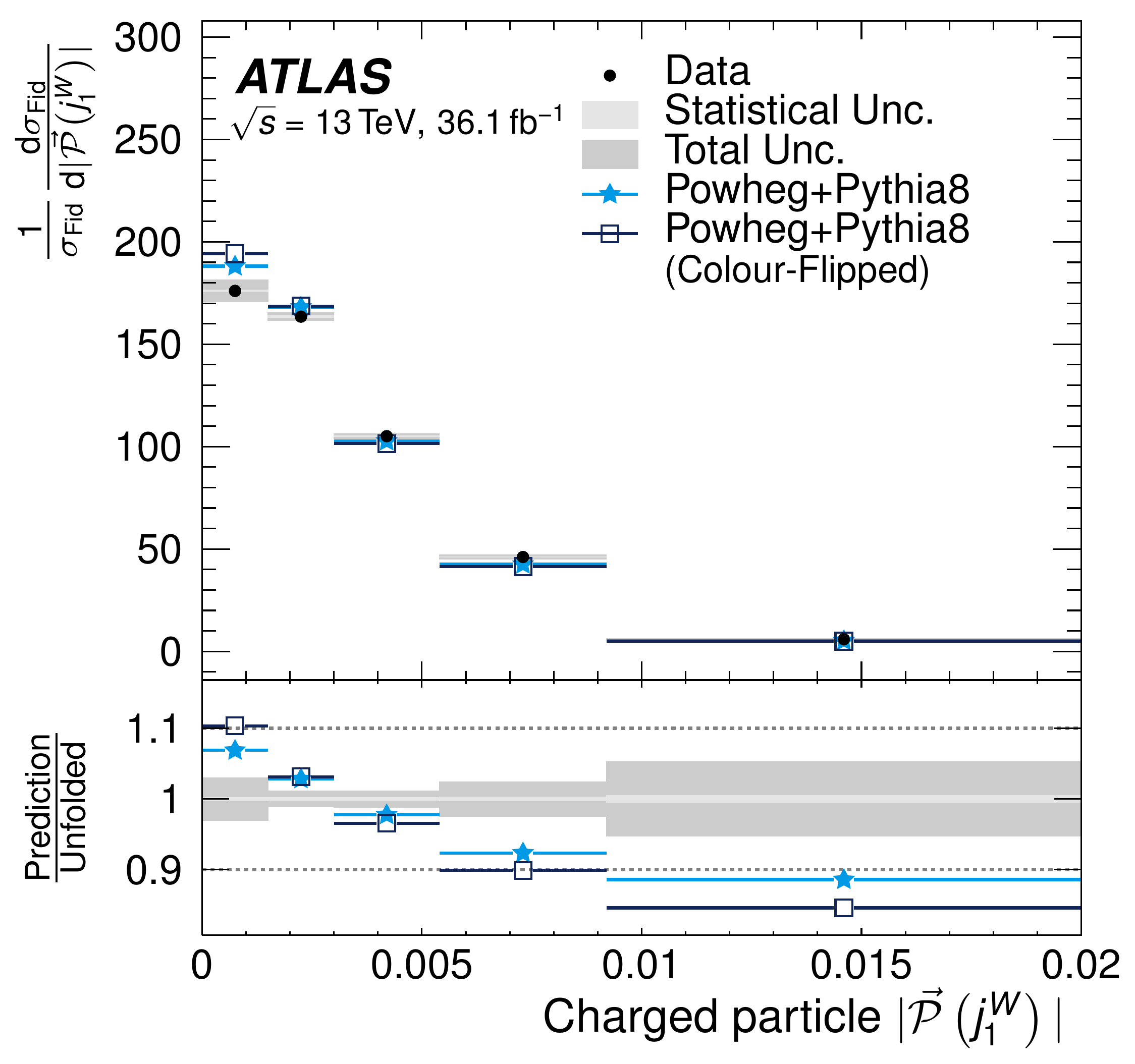}
    \caption{$|\vec{\mathcal{P}}\left( j_1^W \right)|$}
    \label{fig:results-CFlip-NonBFwd-Magnitude}
  \end{subfigure}
  \hfill
  \begin{subfigure}{.49\textwidth}
    \centering
    \includegraphics[width=.99\textwidth]{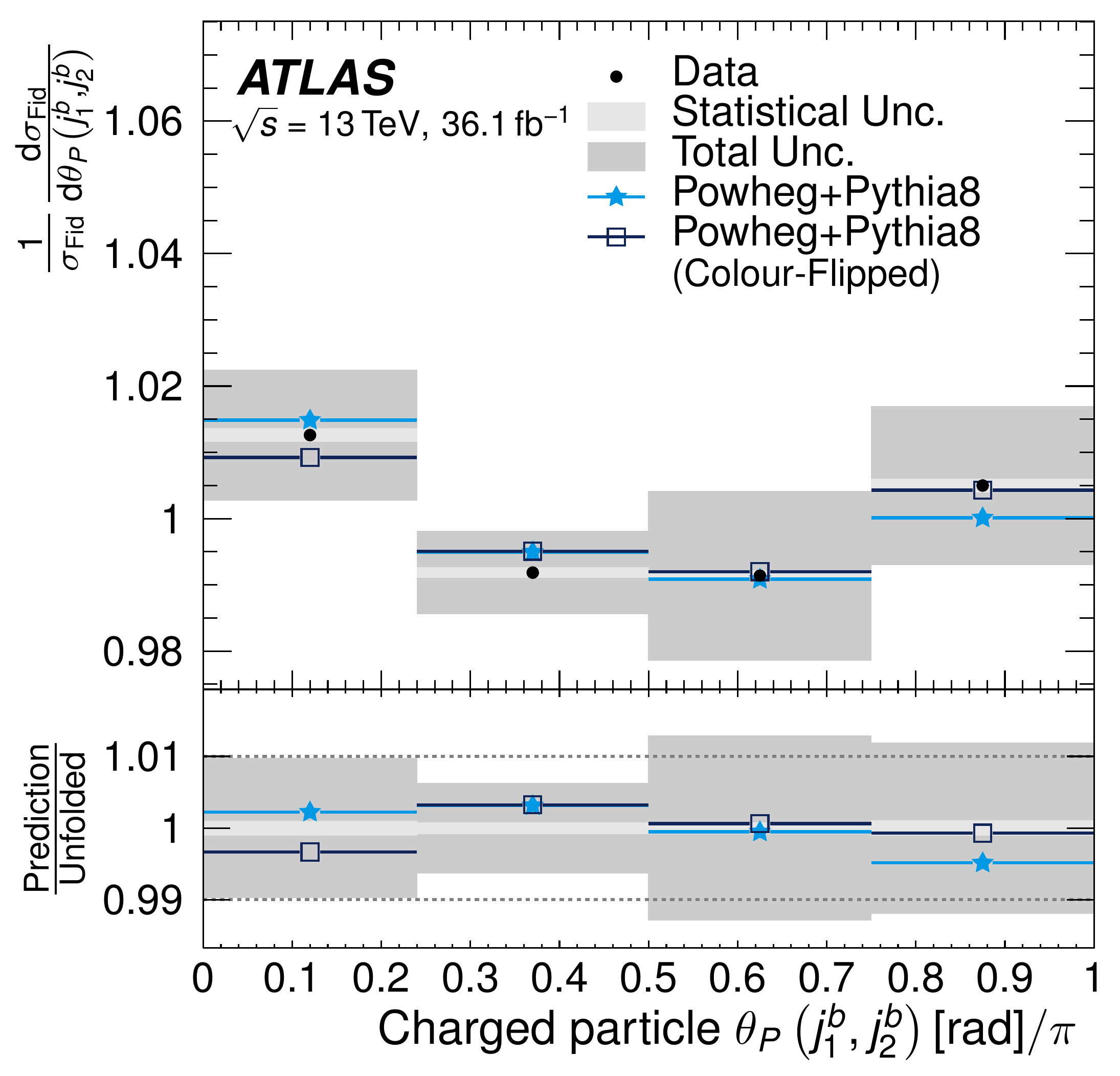}
    \caption{$\theta_{\mathcal{P}}\left( j_1^b, j_2^b \right)$}
    \label{fig:results-CFlip-BFwd-Angle}
  \end{subfigure}
  \caption{
    Normalised fiducial differential cross-sections as a function of the
    \subref{fig:results-CFlip-NonBFwd-Angle}~forward and
    \subref{fig:results-CFlip-NonBBwd-Angle}~backward pull angle for the
    hadronically decaying $W$ boson daughters,
    \subref{fig:results-CFlip-NonBFwd-Magnitude}~the magnitude of the
    leading $W$ daughter's jet-pull vector, and
    \subref{fig:results-CFlip-BFwd-Angle}~the forward di-$b$-jet-pull
    angle.
    The data are compared to a Standard Model prediction
    produced with \POWHEG+~\PYTHIAV{8} as well as the model with exotic colour flow
    also created with \POWHEG+~\PYTHIAV{8}. The uncertainty bands presented in these plots combine the
    baseline set of systematic uncertainties with effects due to considering the
    exotic colour-flipped model as a source of signal modelling
    uncertainty.
    The statistical uncertainties in the predictions are smaller than the marker size.}
  \label{fig:results-CFlip}
\end{figure}

Figure~\ref{fig:results-CFlip} compares the normalised unfolded data to the SM
prediction as well as a prediction obtained from the exotic model with flipped
colour flow described in Section~\ref{sec:data-simulation}.
Both predictions are obtained from MC samples
generated with \POWHEG+~\PYTHIAV{8}. The data agree better with the SM prediction
than the colour-flipped sample.

The uncertainty bands on the unfolding results shown in
Figure~\ref{fig:results-CFlip} include an additional ``colour model
uncertainty''. This uncertainty is obtained using the same procedure that is used for
the signal modelling uncertainties, using the sample with exotic colour flow as
the alternative $t\bar{t}$ MC sample.
It has a similar size to the dominant signal-modelling uncertainties.

A goodness-of-fit procedure is employed in
order to quantify the level of agreement between the measured distributions and
those predicted by the MC generators. A $\chi^2$ test statistic is calculated
for each pairing of an observable and the theoretical prediction individually, using
the full covariance matrix of the experimental uncertainties, but excluding any
uncertainties in the theoretical predictions. Given the unfolded data $D$, the
model prediction $M$, and the covariance $\Sigma$, the $\chi^2$ is given by

\begin{align*}
  % D. Stoker marked this as non-numbered. Previously we've been asked to include numbers even on non-referenced equations.
  \chi^2 = (D^T - M^T) \cdot \Sigma^{-1} \cdot (D - M) \, .
  % \label{eq:chi2}
\end{align*}

Subsequently, $p$-values can be calculated from the $\chi^2$ and number of
degrees of freedom (NDF), and these are the probability to obtain a $\chi^2$ value
greater than or equal to the observed value.

The fact that the analysis measures normalised distributions removes one degree
of freedom from the $\chi^2$ calculation. Consequently, one of the $N$ elements
of $D$ and $M$ is dropped and the covariance is reduced from dimensionality $N
\times N$ to $(N-1) \times (N-1)$ by discarding one column and row. The $\chi^2$
value does not depend on the choice of discarded elements.
Table~\ref{tab:chi2-results} lists the resulting $\chi^2$ values and derived
$p$-values.

\begin{table}[!htb]
  \centering
  \resizebox{.99\textwidth}{!}{\sisetup{
  table-number-alignment=right,
  round-mode=off}
\begin{tabular}{
  l
  S[table-figures-integer=2, table-figures-decimal=1]  @{\hskip+0.0em}r@{\hskip+0.3em}  S[table-figures-integer=1, table-figures-decimal=3]
  S[table-figures-integer=2, table-figures-decimal=1]  @{\hskip+0.0em}r@{\hskip+0.3em}  S[table-figures-integer=1, table-figures-decimal=3]
  S[table-figures-integer=1, table-figures-decimal=1]  @{\hskip+0.0em}r@{\hskip+0.3em}  S[table-figures-integer=1, table-figures-decimal=3]
  S[table-figures-integer=2, table-figures-decimal=1]  @{\hskip+0.0em}r@{\hskip+0.3em}  S[table-figures-integer=1, table-figures-decimal=3]
  }
  \toprule
  \multicolumn{1}{c}{\multirow{2}{*}{Sample}} & \multicolumn{3}{c}{$\theta_{\mathcal{P}} \left( j_1^W, j_2^W \right)$} & \multicolumn{3}{c}{$\theta_{\mathcal{P}} \left( j_2^W, j_1^W \right)$} & \multicolumn{3}{c}{$\theta_{\mathcal{P}} \left( j_1^b, j_2^b \right)$} & \multicolumn{3}{c}{$|\vec{\mathcal{P}}\left( j_1^W \right)|$} \\
  \cmidrule(lr{.75em}){2-4}\cmidrule(lr{.75em}){5-7}\cmidrule(lr{.75em}){8-10}\cmidrule(lr{.75em}){11-13}
  & \multicolumn{2}{c}{$\chi^2 / \text{NDF}$} & \multicolumn{1}{c}{$p$-value} & \multicolumn{2}{c}{$\chi^2 / \text{NDF}$} & \multicolumn{1}{c}{$p$-value} & \multicolumn{2}{c}{$\chi^2 / \text{NDF}$} & \multicolumn{1}{c}{$p$-value} & \multicolumn{2}{c}{$\chi^2 / \text{NDF}$} & \multicolumn{1}{c}{$p$-value} \\
  \midrule
  \text{Powheg+Pythia8} & 50.9 & / 3 & \multicolumn{1}{r}{$< 0.001$} & 25.1 & / 3 & \multicolumn{1}{r}{$< 0.001$} & 0.7 & / 3 & 0.867 & 24.8 & / 4 & \multicolumn{1}{r}{$< 0.001$} \\
  \text{Powheg+Pythia6} & 23.2 & / 3 & \multicolumn{1}{r}{$< 0.001$} & 8.2 & / 3 & 0.042 & 4.2 & / 3 & 0.240 & 21.1 & / 4 & \multicolumn{1}{r}{$< 0.001$} \\
  \text{MG5\_aMC+Pythia8} & 6.8 & / 3 & 0.077 & 6.7 & / 3 & 0.082 & 2.0 & / 3 & 0.563 & 17.6 & / 4 & 0.001 \\
  \text{Powheg+Herwig7} & 2.7 & / 3 & 0.446 & 3.4 & / 3 & 0.328 & 4.8 & / 3 & 0.190 & 11.3 & / 4 & 0.023 \\
  \text{Sherpa} & 22.0 & / 3 & \multicolumn{1}{r}{$< 0.001$} & 11.9 & / 3 & 0.008 & 0.0 & / 3 & 0.998 & 14.1 & / 4 & 0.007 \\
  \midrule
  \text{Powheg+Pythia8}$^{\star}$ & 17.1 & / 3 & \multicolumn{1}{r}{$< 0.001$} & 25.0 & / 3 & \multicolumn{1}{r}{$< 0.001$} & 0.3 & / 3 & 0.958 & 11.1 & / 4 & 0.026 \\
  \text{Flipped Powheg+Pythia8}$^{\star}$ & 45.3 & / 3 & \multicolumn{1}{r}{$< 0.001$} & 45.9 & / 3 & \multicolumn{1}{r}{$< 0.001$} & 2.6 & / 3 & 0.457 & 17.2 & / 4 & 0.002 \\
  \bottomrule
\end{tabular}}
  \caption{The $\chi^2$ and resulting $p$-values for the measured normalised
    cross-sections obtained by comparing the different predictions to the
    unfolded data. When comparing the data with the prediction for the exotic
    flipped colour-flow model, the model itself is considered as an additional
    source of signal modelling uncertainty and thus added to the covariance matrix.
    Calculations that include this additional systematic uncertainty are marked with $\star$.
  }
  \label{tab:chi2-results}
\end{table}

For the signal jet-pull angles $\theta_{\mathcal{P}}\left( j_1^W, j_2^W \right)$ and
$\theta_{\mathcal{P}}\left( j_2^W, j_1^W \right)$, the predictions obtained from
\POWHEG+~\HERWIGV{7} agree best with the observed data. A general trend is that
simulation predicts a steeper distribution, i.e.\ a stronger colour-flow
effect. The magnitude of the jet-pull vector is poorly modelled in general, with the
prediction obtained from \POWHEG+~\HERWIGV{7} agreeing best with data. As with
the signal jet-pull angles, the disagreement shows a similar trend for the
different MC predictions: data favours larger values of the jet-pull vector's
magnitude. Predictions from \POWHEG+~\PYTHIAV{6} are in significantly better
agreement with the data than those obtained from \POWHEG+~\PYTHIAV{8} for the
signal jet-pull angles and jet-pull vector's magnitude.

The signal jet-pull angles and the jet-pull vector's magnitude can be used to
distinguish the case of colour flow like that in the SM from the exotic flipped
colour-flow scenario constructed in Section~\ref{sec:data-simulation}. The data
favour the SM prediction over the colour-flipped prediction.

The forward di-$b$-jet-pull angle is modelled relatively well by most
predictions. In particular the distribution obtained from \SHERPA agrees
extremely well with the measurement. \POWHEG+~\HERWIGV{7}, which otherwise shows
relatively good agreement with data for the other three observables, agrees
least well of the tested predictions. Indeed, it is the only prediction that is
consistently outside of the estimated uncertainty bands. As expected, the
forward di-$b$-jet-pull angle $\theta_{\mathcal{P}}\left(j_1^b, j_2^b\right)$
does not show the sloped distribution that the signal jet-pull angles
$\theta_{\mathcal{P}}\left( j_1^W, j_2^W\right)$ and $\theta_{\mathcal{P}}\left(
j_2^W, j_1^W\right)$ follow.

% -------------------------------------------------------------------------------
\section{Conclusion}
\label{sec:conclusion}
% -------------------------------------------------------------------------------

A measurement of four observables sensitive to the colour flow in $t\bar{t}$
events is presented, using $36.1\,\text{fb}^{-1}$ of $\sqrt{s} = \SI{13}{\TeV}$
$pp$ collision data recorded by the ATLAS detector at the LHC. The four
observables are the forward and backward jet-pull angles for the $W$ boson
daughters, the magnitude of the jet-pull vector of the leading $W$ boson
daughter, and the jet-pull angle between the $b$-tagged jets. The measured distributions are compared to several theoretical predictions
obtained from MC simulation.

The default SM prediction, \POWHEG+~\PYTHIAV{8}, agrees poorly with the data.
However, alternative SM predictions exhibit much better agreement. In
particular, the prediction obtained by \POWHEG+~\HERWIGV{7} provides a rather
good description of the data. Predictions from \POWHEG+~\PYTHIAV{6} are in
significantly better agreement with the data than those obtained from
\POWHEG+~\PYTHIAV{8}.

In addition, a model with exotic colour flow is compared to the data. In the
observables sensitive to the exotic colour flow, data favours the SM case over
the exotic model.

\makeatletter
\ifADOC@PAPER

\section*{Acknowledgements}
% Acknowledgements for papers with collision data
% Version 14-Feb-2018

% Standard acknowledgements start here
%----------------------------------------------
We thank CERN for the very successful operation of the LHC, as well as the
support staff from our institutions without whom ATLAS could not be
operated efficiently.

We acknowledge the support of ANPCyT, Argentina; YerPhI, Armenia; ARC, Australia; BMWFW and FWF, Austria; ANAS, Azerbaijan; SSTC, Belarus; CNPq and FAPESP, Brazil; NSERC, NRC and CFI, Canada; CERN; CONICYT, Chile; CAS, MOST and NSFC, China; COLCIENCIAS, Colombia; MSMT CR, MPO CR and VSC CR, Czech Republic; DNRF and DNSRC, Denmark; IN2P3-CNRS, CEA-DRF/IRFU, France; SRNSFG, Georgia; BMBF, HGF, and MPG, Germany; GSRT, Greece; RGC, Hong Kong SAR, China; ISF, I-CORE and Benoziyo Center, Israel; INFN, Italy; MEXT and JSPS, Japan; CNRST, Morocco; NWO, Netherlands; RCN, Norway; MNiSW and NCN, Poland; FCT, Portugal; MNE/IFA, Romania; MES of Russia and NRC KI, Russian Federation; JINR; MESTD, Serbia; MSSR, Slovakia; ARRS and MIZ\v{S}, Slovenia; DST/NRF, South Africa; MINECO, Spain; SRC and Wallenberg Foundation, Sweden; SERI, SNSF and Cantons of Bern and Geneva, Switzerland; MOST, Taiwan; TAEK, Turkey; STFC, United Kingdom; DOE and NSF, United States of America. In addition, individual groups and members have received support from BCKDF, the Canada Council, CANARIE, CRC, Compute Canada, FQRNT, and the Ontario Innovation Trust, Canada; EPLANET, ERC, ERDF, FP7, Horizon 2020 and Marie Sk{\l}odowska-Curie Actions, European Union; Investissements d'Avenir Labex and Idex, ANR, R{\'e}gion Auvergne and Fondation Partager le Savoir, France; DFG and AvH Foundation, Germany; Herakleitos, Thales and Aristeia programmes co-financed by EU-ESF and the Greek NSRF; BSF, GIF and Minerva, Israel; BRF, Norway; CERCA Programme Generalitat de Catalunya, Generalitat Valenciana, Spain; the Royal Society and Leverhulme Trust, United Kingdom.

The crucial computing support from all WLCG partners is acknowledged gratefully, in particular from CERN, the ATLAS Tier-1 facilities at TRIUMF (Canada), NDGF (Denmark, Norway, Sweden), CC-IN2P3 (France), KIT/GridKA (Germany), INFN-CNAF (Italy), NL-T1 (Netherlands), PIC (Spain), ASGC (Taiwan), RAL (UK) and BNL (USA), the Tier-2 facilities worldwide and large non-WLCG resource providers. Major contributors of computing resources are listed in Ref.~\cite{ATL-GEN-PUB-2016-002}.
%----------------------------------------------

\fi
\makeatother

%-------------------------------------------------------------------------------
\clearpage
\appendix
% \part*{Appendix}
% \addcontentsline{toc}{part}{Appendix}
%-------------------------------------------------------------------------------

% -------------------------------------------------------------------------------
% If you use biblatex and either biber or bibtex to process the bibliography
% just say \printbibliography here
\printbibliography
% If you want to use the traditional BibTeX you need to use the syntax below.
% \bibliographystyle{bibtex/bst/atlasBibStyleWoTitle}
% \bibliography{colour_flow,bibtex/bib/ATLAS}
% -------------------------------------------------------------------------------

% -------------------------------------------------------------------------------
% Print the list of contributors to the analysis
% The argument gives the fraction of the text width used for the names
% -------------------------------------------------------------------------------
\clearpage
% ATLAS Collaboration author list
% Reference date of TOPQ-2017-13 is 2017-09-17
% Author list last updated on date 07-MAY-18
% Data extracted on 07-May-2018 for paper reference TOPQ-2017-13
% at 6:07pm
 
\begin{flushleft}
{\Large The ATLAS Collaboration}

\bigskip

M.~Aaboud$^\textrm{\scriptsize 34d}$,
G.~Aad$^\textrm{\scriptsize 99}$,
B.~Abbott$^\textrm{\scriptsize 124}$,
O.~Abdinov$^\textrm{\scriptsize 13,*}$,
B.~Abeloos$^\textrm{\scriptsize 128}$,
S.H.~Abidi$^\textrm{\scriptsize 164}$,
O.S.~AbouZeid$^\textrm{\scriptsize 143}$,
N.L.~Abraham$^\textrm{\scriptsize 153}$,
H.~Abramowicz$^\textrm{\scriptsize 158}$,
H.~Abreu$^\textrm{\scriptsize 157}$
% ...
\bigskip
\\

% ...
$^{13}$Institute of Physics, Azerbaijan Academy of Sciences, Baku; Azerbaijan.\\
% ...
$^{34}$$^{(a)}$Facult\'e des Sciences Ain Chock, R\'eseau Universitaire de Physique des Hautes Energies - Universit\'e Hassan II, Casablanca;$^{(b)}$Centre National de l'Energie des Sciences Techniques Nucleaires, Rabat;$^{(c)}$Facult\'e des Sciences Semlalia, Universit\'e Cadi Ayyad, LPHEA-Marrakech;$^{(d)}$Facult\'e des Sciences, Universit\'e Mohamed Premier and LPTPM, Oujda;$^{(e)}$Facult\'e des sciences, Universit\'e Mohammed V, Rabat; Morocco.\\
% ...
$^{99}$CPPM, Aix-Marseille Universit\'e and CNRS/IN2P3, Marseille; France.\\
% ...
$^{124}$Homer L. Dodge Department of Physics and Astronomy, University of Oklahoma, Norman OK; United States of America.\\
% ...
$^{128}$LAL, Universit\'e Paris-Sud, CNRS/IN2P3, Universit\'e Paris-Saclay, Orsay; France.\\
% ...
$^{143}$Santa Cruz Institute for Particle Physics, University of California Santa Cruz, Santa Cruz CA; United States of America.\\
% ...
$^{153}$Department of Physics and Astronomy, University of Sussex, Brighton; United Kingdom.\\
% ...
$^{157}$Department of Physics, Technion: Israel Institute of Technology, Haifa; Israel.\\
$^{158}$Raymond and Beverly Sackler School of Physics and Astronomy, Tel Aviv University, Tel Aviv; Israel.\\
% ...
$^{164}$Department of Physics, University of Toronto, Toronto ON; Canada.\\
% ...
$^{a}$ Also at Borough of Manhattan Community College, City University of New York, New York City; United States of America.\\
$^{b}$ Also at Centre for High Performance Computing, CSIR Campus, Rosebank, Cape Town; South Africa.\\
$^{c}$ Also at CERN, Geneva; Switzerland.\\
$^{d}$ Also at CPPM, Aix-Marseille Universit\'e and CNRS/IN2P3, Marseille; France.\\
$^{e}$ Also at Departament de Fisica de la Universitat Autonoma de Barcelona, Barcelona; Spain.\\
% ...
$^{*}$ Deceased

\end{flushleft}

% Created with Glance <Atlas.Glance@cern.ch>

% \PrintAtlasContribute{0.30}
% -------------------------------------------------------------------------------

%-------------------------------------------------------------------------------
% Auxiliary material - comment out the following line if you do not have any
% \include{colour_flow-auxmat}
%-------------------------------------------------------------------------------

\end{document}